\documentclass[5p,times]{elsarticle}

\usepackage{lineno,hyperref}
\modulolinenumbers[5]

\journal{ }
\usepackage[british]{babel}

\biboptions{authoryear}

\usepackage{amsmath,amssymb,amsfonts}
\usepackage{graphicx}
\usepackage{textcomp}
\usepackage[table]{xcolor}

\usepackage{algorithm}
\usepackage{algpseudocode}
\usepackage{booktabs}
\usepackage{lineno}

\usepackage{subfig}
\usepackage{bm}
\usepackage{caption}

\hypersetup{%
  colorlinks=true,
}

\def\BibTeX{{\rm B\kern-.05em{\sc i\kern-.025em b}\kern-.08em
    T\kern-.1667em\lower.7ex\hbox{E}\kern-.125emX}}

\usepackage{cleveref}

\begin{document}
\let\today\relax

\begin{frontmatter}

\title{A novel dynamic asset allocation system using Feature Saliency  Hidden Markov models for smart beta investing}


\author[AB,Manchester]{Elizabeth Fons\corref{mycorrespondingauthor}}
\ead[email]{elizabeth.fons@postgrad.manchester.ac.uk}

\author[AB]{Paula Dawson}

\author[AB,Berkeley]{Jeffrey Yau}

\author[Manchester]{Xiao-jun Zeng}

\author[Manchester]{John Keane}

\address[AB]{AllianceBernstein, London, UK.}
\address[Berkeley]{University of California, Berkeley, California, USA.}
\address[Manchester]{School of Computer Science, University of Manchester, Manchester, UK.}


\begin{abstract}
The financial crisis of 2008 generated interest in more transparent, rules-based strategies for portfolio construction, with Smart beta strategies emerging as a trend among institutional investors. While they perform well in the long run, these strategies often suffer from severe short-term drawdown (peak-to-trough decline) with fluctuating performance across cycles. To address cyclicality and underperformance, we build a dynamic asset allocation system using Hidden Markov Models (HMMs). We test our system across multiple combinations of smart beta strategies and the resulting portfolios show an improvement in risk-adjusted returns, especially on more return oriented portfolios (up to 50$\%$ in excess of market annually). In addition, we propose a novel smart beta allocation system based on the Feature Saliency HMM (FSHMM) algorithm that performs feature selection simultaneously with the training of the HMM, to improve regime identification. We evaluate our systematic trading system with real life assets using MSCI indices; further, the results (up to 60$\%$ in excess of market annually) show model performance improvement with respect to portfolios built using full feature HMMs .\\

\end{abstract}


\end{frontmatter}

\section{Introduction}

Smart beta is a relatively new term that has become ubiquitous in asset management over the last few years. The financial theory underpinning Smart Beta, known as factor investing, has been around since the 1960s, when factors were first identified as being drivers of equity returns \citep{Agather2017}. These factor returns can be a source of risk and/or improved return, and understanding whether any additional risk is adequately compensated with higher returns is important. \citep{Ang:2014}.

By selecting stocks based on their factor exposures, active managers can build portfolios with particular factor exposures and so use factor investing to improve portfolio returns and/or lower risk, depending on their particular objectives.  Smart beta aims to achieve these goals at a reduced cost by utilising a transparent, systematic, rules-based approach, bringing down the costs significantly when compared to active management \citep{Asness2016}.

While smart beta strategies have shown strong performance in the long run, they  often suffer from severe short-term drawdown (peak-to-trough decline) with fluctuating performance across cycles \citep{Arnott2016}. These fluctuations can arise from extreme macroeconomic conditions, elevated volatility, heightened correlations across multiple markets and uncertainty monetary and fiscal policy responses. 
In this paper we address this by building a regime switching model using Hidden Markov Models (HMMs). 
Hidden Markov models have become one of the mainstream techniques to model times series data \citep{baum1970, Rabiner:1989}, with applications across many areas such as speech recognition, text classification and medical applications. 
We first study how a regime switching framework can be used to detect regimes across factors and, if so, add value to smart beta strategies. The prevalent approach in regime switching frameworks for asset allocation has been to specify in advance a static decision rule dependent on the predicted state \citep{Nystrup:2018}. 
An alternative approach is to dynamically optimise a portfolio using information from the inferred regime parameters. We follow this second approach and use the regime information to construct different types of portfolios (more return oriented and more risk focused). In a first step we build a dynamic asset allocation (DAA) system to construct portfolios through a regime switching model and perform a systematic analysis using hundreds of combinations of factors by  training the HMM with the same factors that will be used for the allocation in the portfolio. Our study shows that using the regime information from the HMM has a better performance than a single regime allocation and we find that more return-oriented portfolios yield better risk-adjusted returns than their benchmarks, while the performance of more risk focused portfolios show some improvement. 

Finally,  the common factor in the majority of the research on regime-switching models in finance is that it considers either a single or a small set of assets to build the model, with the selection criteria for the assets usually coming from domain knowledge. The reason for this is that unsupervised feature selection for HMMs is very limited, with wrapping methods exhibiting high computational cost or with very few methods specific for HMMs \citep{FSHMMsSurvey}. In most applications of HMMs, features are either pre-selected based on expert knowledge or feature selection is omitted entirely.  One of the few feature selection algorithms developed for HMMs is the feature saliency hidden Markov model (FSHMM) proposed by \citet{FSHMM:article}, where the feature selection process is embedded in the training of the HMM. We incorporate this FSHMM into our dynamic asset allocation system. with two benefits: (1) by selecting the features during the training we expect to improve regime identification by selecting features that are state dependent and rejecting features that are state independent; (2) it allows incorporation of many features on a model and let the algorithm decide which ones contribute to regime identification, thus avoiding the need for expert knowledge in the construction of financial cycles.

The main contributions of this paper are the following:

\begin{enumerate}
    \item We build a dynamic asset allocation (DAA) system using an HMM for regime detection and perform a systematic study using multiple combinations of assets and comparing performance with their single-regime portfolio counterparts. We show that the DAA system consistently performs better than the benchmarks;
    \item We extend our DAA system by incorporating a Feature Saliency HMM for feature selection, thus improving regime identification;
    \item We test the DAA system with embedded feature selection on real life investable indices using MSCI indices and show an improvement in risk-adjusted return on strategies built using the DAA system with FSHMM with respect to strategies built using DAA system without feature selection. 
\end{enumerate}

This paper is organized as follows: 
Section \ref{section:prev_work} gives an overview of previous work on HMM in finance; Section \ref{section:RSframework} introduces hidden Markov models and feature saliency hidden Markov models; data and index construction are described in Section \ref{section:data_and_performance}; 
Section \ref{section:DAA_system} introduces the dynamic allocation system, the feature saliency algorithm and its incorporation into our dynamic asset allocation system; Section \ref{section:results} shows the experimental results of the DAA system, and the incorporation of embedded feature selection. Finally, we test the DAA system with feature selection using investable assets; conclusions and further work are considered  Section \ref{section:conclusion}. 

\section{Previous work}
\label{section:prev_work}

In finance, HMMs have been used extensively  to build regime-based models, since Hamilton proposed using a regime-switching model to identify economic cycles using the GNP series \citep{Hamilton:1989}. As pointed out by \citet{Ang2012} HMMs can simultaneously capture multiple characteristics from financial return series such as time-varying correlations, skewness and kurtosis, while also providing good approximations even in processes for which the underlying model is unknown \citep{Ang:2004, Bulla:2011a, Bulla:2006, Nystrup:2015, Nystrup:2017}. In addition, HMMs allow for good interpretability of results, as thinking in terms of regimes is a natural approach in finance. Examples of dynamic asset allocation are \citet{ReusMulvey:2016} that use a HMM to build a dynamic portfolio using currency futures and \citet{BaeMulvey:2014} that use a HMM to identify market regimes using different asset classes, with regime information helping portfolios to avoid risk during left-tail events.

\citet{Guidolin2012} provides an extensive review on applications of Markov switching models in empirical finance covering stock returns, term structure of default-free interest rates, exchange rates and joint processes of stock and bond returns.  

Outside of asset allocation, HMMs have been used to capture energy prices dynamics \citep{Ramos:2014} to build credit risk systems, for example \citet{Petropoulos:2016} build a credit rating system using a students'-t HMM, addressing two problems in current systems, their heavy-tailed actual distribution and their time-series nature; \citet{Elliott:2014} build a model using double hidden Markov model to extract information about true credit qualities of firms.
\citet{Dabrowski:2016} study HMMs and other Bayesian networks to build early warning systems to detect systemic banking crisis and find that Bayesian methods provided  superior performance on early warning than traditional signal extraction logic models and \citet{Zhou:2012} investigate three popular short-rate models and extend them to capture the switching of economic regimes using a finite-state Markov chain.

So far, little work has been done on the impact of regime switching models to factor investing.  Among them, \citet{Guidolin2008} found evidence of four economic regimes in size and value factors that capture time-variations in mean returns, volatilities and return correlations.  \citet{Zhao:2011a} and \citet{Zhao:2011b}  study time-varying risk premiums using a six factor model to explain the returns of sector ETFs. In their work they cover a short period of testing time (9 months) and do not consider transaction costs.

\section{Theoretical background}
\label{section:RSframework}

In this section we present the hidden Markov model and the feature saliency hidden Markov model that can simultaneously train the model and perform feature selection.

\subsection{Hidden Markov Models (HMMs)}
HMMs are sequential models that assume an underlying hidden process modeled by a Markov chain and a sequence of observed data as a noisy manifestation of this latent process \citep{Murphy:2012}. 

\begin{figure}[h!]
\centering
  \includegraphics[width=0.4\textwidth]{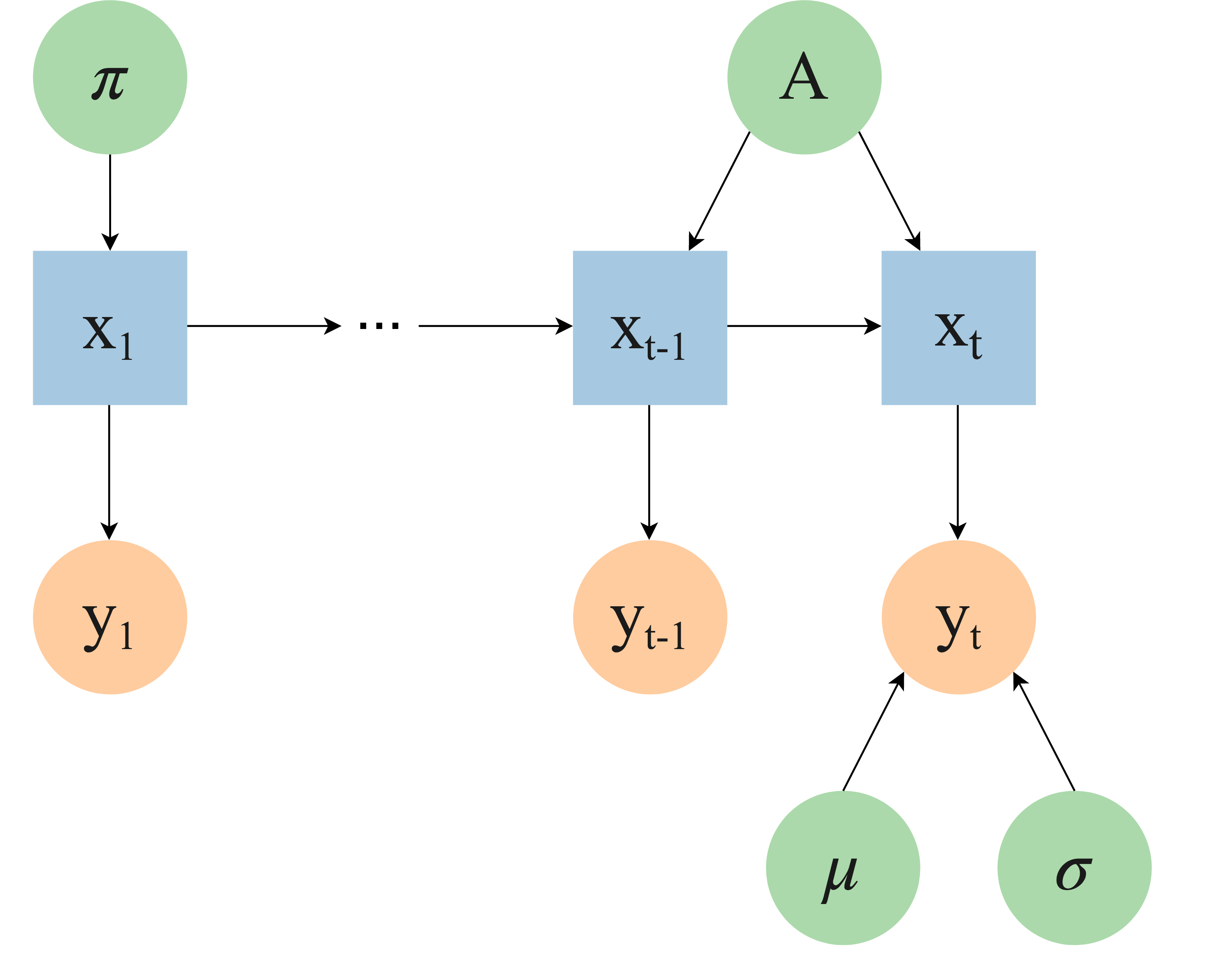}
  \caption{The Hidden Markov Model: blue squares represent latent variables, orange circles are  observations and green circles represent model parameters.}
  \label{fig:hmm}
\end{figure}

Given $o = \{\mathbf{y_1}, ... , \mathbf{y_T}\}$ the sequence of observed data where each $x_t \in \mathbb{R}^L$ with $L$ the dimension of observations and $x = {x_1, \ldots, x_T}$ the latent sequence of states where $x_t \in \{1, \ldots, K\}$ with $K$ the number of latent states.
The HMM model parameters are $\Lambda = (\pi, A, \mu, \sigma)$ where $\pi$ and $A$ correspond to the initial probability and transition probabilities, and $\mu$ and $\sigma$ are the mean and variance of the state dependent Gaussian feature distribution (generally called emission probabilities, symbolized here by $b_{x_t}$), the graphical model of the HMM can be seen in Figure \ref{fig:hmm} where blue squares represent latent variables, orange circles are observations and green circles represent model parameters. The complete likelihood can be written as:.  

\begin{equation}
    p(x,y|\Lambda) = \pi(x_0) b_{x_0}(y_0) \prod^T_{t=1} A(x_{t-1},x_t) b_{x_t}(y_t)
    \label{eq:HMMlike}
\end{equation}


In this work the sequence of noisy observations are factor indices returns and the underlying hidden process is the state of the market that generates them. We assume that the emission probabilities are Gaussian. While normal distributions are a poor fit to financial returns, the mixture of normal distributions provide a much better fit capturing stylize behaviors including fat tails and skewness \citep{Nystrup:2015, Ang2012}. 


The training of HMMs is done by the Baum-Welch algorithm, a type of Expectation-Maximization (EM) algorithm \citep{Rabiner:1989}. The E-step calculates the expected value of the log-likelihood with respect to the state, given the data and current model parameters and the M-step maximizes the expectation computed in the previous step to update the model parameters. The algorithm iterates between these two steps until convergence. The expectation of the complete log-likelihood function is given by:
\begin{equation}
    Q(\Lambda, \Lambda') = E[\log{p(x,y|\Lambda)}|y,\Lambda']
    \label{eq:qML}
\end{equation}
where $\Lambda$ are the parameters for the current iteration and $\Lambda^{\prime}$ is the set of parameters from the previous iteration. 

Following \citet{FSHMM:article}, we place priors on the parameters and calculate the MAP estimate, so the $Q$ function is modified by adding the prior on the model parameters, $G(\Lambda)$:
\begin{equation}
     Q(\Lambda, \Lambda') + \log{G(\Lambda)}
     \label{eq:qMAP}
\end{equation}
The EM algorithm is as follows, the $Q$ function in \ref{eq:qML} is calculated in the E-step and the equation \ref{eq:qMAP} is maximized in the M-step.

\subsection{FSHMM}
\label{section:FSHMMtheory}

\begin{figure}[h!]
\centering
  \includegraphics[width=0.4\textwidth]{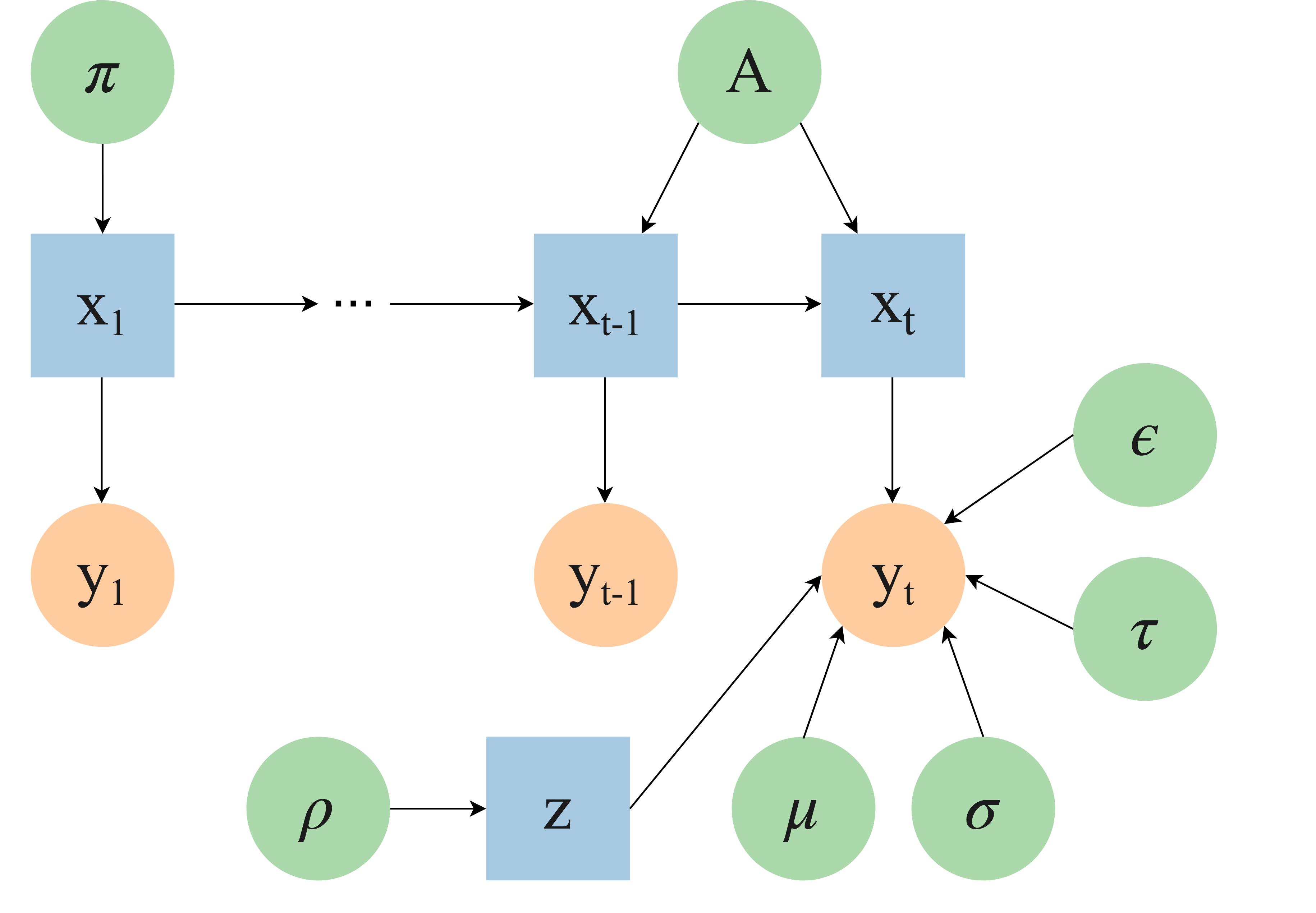}
  \caption{The feature saliency Hidden Markov Model: blue squares represent latent variables, orange circles are observations and green circles represent model parameters.}
  \label{fig:FSHMM}
\end{figure}

The feature saliency HMM considers a feature relevant if its distribution is dependent on the underlying state and irrelevant if it is independent. Given a set of binary variables $\{z_1, \ldots, z_L\}$ that indicate the relevance of the feature, i.e. $z_l = 1$ if the $l$-th feature is relevant and $z_l =0$ if it's irrelevant, the feature saliency $\rho_l$ is defined as the probability that the $l$-th feature is relevant. 
Assuming the features are conditionally independent given the state enables the multivariate Gaussian to be written as a multiplication of univariate Gaussians, and the conditional distribution of $y_t$ given $z$ and $x$ can be written as follows:
\begin{equation}
    p(y_t|z,x_t=i, \Lambda) = \prod_{l=1}^L r(y_{lt}|\mu_{il},\sigma^2_{il})^{z_l} q(y_{lt}|\epsilon_l,\tau^2_l)^{1-z_l}
\end{equation}
where $r(y_{lt}|\mu_{il},\sigma^2_{il})$ is the Gaussian conditional feature distribution for the $l$-th feature and $q(y_{lt}|\epsilon_l,\tau^2_l)$ is the state-independent feature distribution.
The FSHMM model parameters are $\Lambda = (\pi, A, \mu, \sigma, \rho, \epsilon, \tau)$ where the first four parameters correspond to the regular HMM, $\rho$ is the feature saliency and $\epsilon$ and $\tau$ are the mean and variance of the state independent Gaussian feature distribution. Figure \ref{fig:FSHMM} shows the feature saliency Hidden Markov Model. 

The marginal probability of $z$ is:
\begin{equation}
    p(z|\Lambda) = \prod_{l=1}^L \rho_l^{z_l} (1-\rho_l)^{1-z_l}
\end{equation}
The joint probability distribution of $y_t$ and $z$ given $x$ is:
\begin{align}
    & p(y_t,z|x_t=i,\Lambda) = \nonumber \\
    & \prod_{l=1}^L [\rho_l r(y_{lt}|\mu_{il},\sigma^2_{il})]^{z_l} [(1-\rho_l)q(y_{lt}|\epsilon_l,\tau^2_l)]^{1-z_l}
\end{align}

The complete likelihood for the FSHMM is given by:
\begin{equation}
    p(x,y,z|\Lambda) = \pi_{x_0} p(y_0,z|x_0,\Lambda) \prod_{t=1}^L a_{x_{t-1},x_t} p(y_t,z|x_t,\Lambda)
\end{equation}

The MAP estimation of the FSHMM is similar to the HMM using EM but the $Q$ function incorporates the hidden variables associated with feature saliency and can be written as:

\begin{align}
    Q(\Lambda,\Lambda^{\prime}) & = E[\log{p(x,y,z|\Lambda)}|y,\Lambda^{\prime}]\nonumber \\
    & = \sum_{x,z} \log{p(x,y,z|\Lambda)} P(x,z|y,\Lambda^{\prime})
\end{align}

The update steps of the EM algorithm are shown in Appendix \ref{appendix:FS} and the pseudocode for the MAP FSHMM formulation is given in Algorithm \ref{fhsmm-algo}. A detailed description of the equation derivations and the steps of the algorithm can be found in \citet{Adams2015}. 

\begin{algorithm}
\caption{MAP FSHMM Algorithm}\label{fhsmm-algo}
\begin{algorithmic}[1]
\State Select initial values for $\pi_i, a_{ij}, \mu_{il}, \sigma_{il}, \epsilon_l, \tau_l$ and $\rho_l$ for $i=1\ldots I, j = 1 \ldots I$, and $l=1\ldots L$
\State Select initial values for $\bar{p}_i, \bar{a}_{ij}, m_{il}, s_{il}, \zeta_{il}, \eta_{il}, b_l, c_l, \nu_l, \psi_l$ and $k_l$ for $i=1\ldots I, j = 1 \ldots I$, and $l=1\ldots L$
\State Select stopping threshold $\delta$ and maximum number of iterations $M$
\State Set absolute percent change in the posterior probability between current iteration and previous iteration $\Delta \mathcal{L}$ to $\infty$ and the number of iterations $it$ to 1
\While {$\Delta \mathcal{L} > \delta$ and $it < M$}
  \State E-step: calculate probabilities $\gamma_t(i), \xi(i,j), e_{ilt}, h_{ilt}, g_{ilt},$ $u_{ilt}, v_{ilt}$ following \ref{eq:Estep01} to \ref{eq:Estep5} 
   \State M-step: update parameters $\pi_i,a_{ij},\mu_{il},\sigma^2_{il}, \epsilon_l, \tau^2_l, \rho_l$ following \ref{eq:Mstepi} to \ref{eq:Mstepf}
  \State $\Delta \mathcal{L}$
  \State $it = it+1$
\EndWhile
\State Perform feature selection based on $\rho_l$ and construct reduced models
\end{algorithmic}
\end{algorithm}

As well as the parameters estimated through EM, the model also has several hyperparameters to set in advance. The most relevant is the weight parameter $k_l$ that can be used as an informative exponential prior on $\rho$. Setting higher values of $k_l$ for the parameters translates into a higher cost in the algorithm, so in order for the algorithm to select that feature, it needs more evidence that this feature is relevant. This can either be used to reduce the number of selected features or as a proxy for the cost of selecting a feature in the optimization process. The heuristic to select a reasonable value of $k_l$ is to scale it with the number of observations as $T/4$ with $T$ the number of observations.

\subsection{Smart Beta investing}
\label{subsection:smartBeta}

As mentioned, smart beta is a systematic, low cost implementation of factor investing, where securities are selected based on their exposure to an attribute that has been associated with a persistent higher return in the past, called a factor.   Factors can be fundamental characteristics of the economy (macroeconomic factors) or of companies (style factors).  Macroeconomic factors can be thought of as capturing the broad risks and returns across assets classes while style factors can be thought of as aiming to explain returns and risks for securities within asset classes.
 
This paper looks at style factors in the equity market. Within style factors, dozens of indicators have been identified. The majority can be grouped into families, with style factors within a family measuring similar characteristics and often highly correlated.  An example of this is momentum, which includes factors measuring returns over different periods (3-months, 6-months, 12-months etc).  While there is no universal definition of these families or the factors that belong in each family there are common themes.  Typically, families will comprise: value, growth, momentum, quality, size and some sort of volatility/risk/beta measure.  There may be variations on this, for example Dividend Yield is sometimes viewed as a factor family in its own right or sometimes it is viewed as a member of the Value family; sometimes the Value family can be split into Value and Deep Value.

\section{Data}
\label{section:data_and_performance}

Below is the description of the two datasets used, and table \ref{table:datasets} summarises their main characteristics.

\subsubsection*{Daily factor data from S$\&$P500 index}

The first dataset is a set of style factors which are constructed based on the S$\&$P 500 universe of US stocks.  The style factor for each individual stock is determined, the universe is ranked and a portfolio is constructed with the top 20$\%$ of stocks and short positions (negative weights) in the bottom 20$\%$ of stocks.  This is repeated each month. The resulting style factor portfolio will have a strong exposure to the factor and no exposure to the overall market (because the negative holdings offset the positive weights) - Table \ref{table:JPM} shows these.  The data is supplied by a broker and consists of 25 style factors covering a time period from 1988 to 2016. This dataset is used throughout the analysis.

\newcommand{\ra}[1]{\renewcommand{\arraystretch}{#1}}
\begin{table*}[t]
	\caption{Representative factor indices used for building regime switching frameworks.}
	\ra{1.2}
	\centering
	\footnotesize
	\begin{tabular}{@{}ccccccc@{}}\toprule
		$\#$ & Factor & Family & & $\#$  & Factor & Family \\
		\cmidrule{1-3}  \cmidrule{5-7}
		1 & Book Value Yield & Value &  & 14 & Operating Margin Growth-1Yr & Quality\\
		2 & 1 Yr Fwd Earnings Yield & Value & & 15 & Operating Margin Growth-3Yr & Quality\\
        3 & Free Cash Flow Yield & Value & & 16 & Historical Free Cash Flow Growth-1Yr & Growth \\
        4 & Sales Yield & Value & & 17 & Historical Free Cash Flow Growth-3Yr & Growth \\
        5 & Dividend Yield & Value & & 18 & Historical DPS Growth-1Yr & Growth\\
        6 & Historical ROE & Quality & & 19 & Historical DPS Growth-3Yr & Growth\\
		7 & Operating (EBIT) Margin & Quality & & 20  & 6 Month Price Momentum & Momentum \\
		8 & AltmanZ & Quality & & 21 & 12 Month Price Momentum & Momentum\\
        9 & ROA & Quality & & 22 & 3 Month Avg Mean EPS & Quality\\
        10 & Piotroski & Quality & & 23 & Size & Risk\\
        11 & Earnings Growth FY1 to FY2 & Growth & & 24 & EPSCV & Quality \\
        12 & Historical Sales Growth-1Yr & Growth & & 25 & Beta & Risk\\
        13 & Historical Sales Growth-3Yr & Growth & & & &\\
		\bottomrule
	\end{tabular}
	\label{table:JPM}
\end{table*}

\subsubsection*{Daily MSCI USA enhanced indices}

The second dataset is supplied by MSCI and consists of a range of indices which they publish.  Like the first dataset, the individual style factors are calculated using underlying stocks and their style factor exposures.  These individual style factor indices are then grouped into six style factor families, and it’s these indices that are used in this paper.  We use the six MSCI USA enhanced style indices, which are: value, low size, momentum, quality, low volatility and dividend yield \cite{MSCI:tablecitation}.  These have different inception dates, with the most recent beginning in 1999, which limits the period we can use this dataset for to 1999-2016.

The advantage of using a published set of indices (such as the MSCI indices) is that they can be packaged into an easy to purchase product, such as an Exchange Traded Fund (ETF), by a separate investment company.  As an example, an investor who wants to buy US value stocks can buy an MSCI US enhanced Value ETF, which would involve buying one security (the ETF) rather than the underlying stocks. By removing the need to analyse and purchase the underlying companies, the complexity and cost of implementing a smart beta strategy can be reduced.  This allows us to test our Novel DAA system with real world assets.

\begin{figure}[h!]
\centering
  \includegraphics[width=0.5\textwidth]{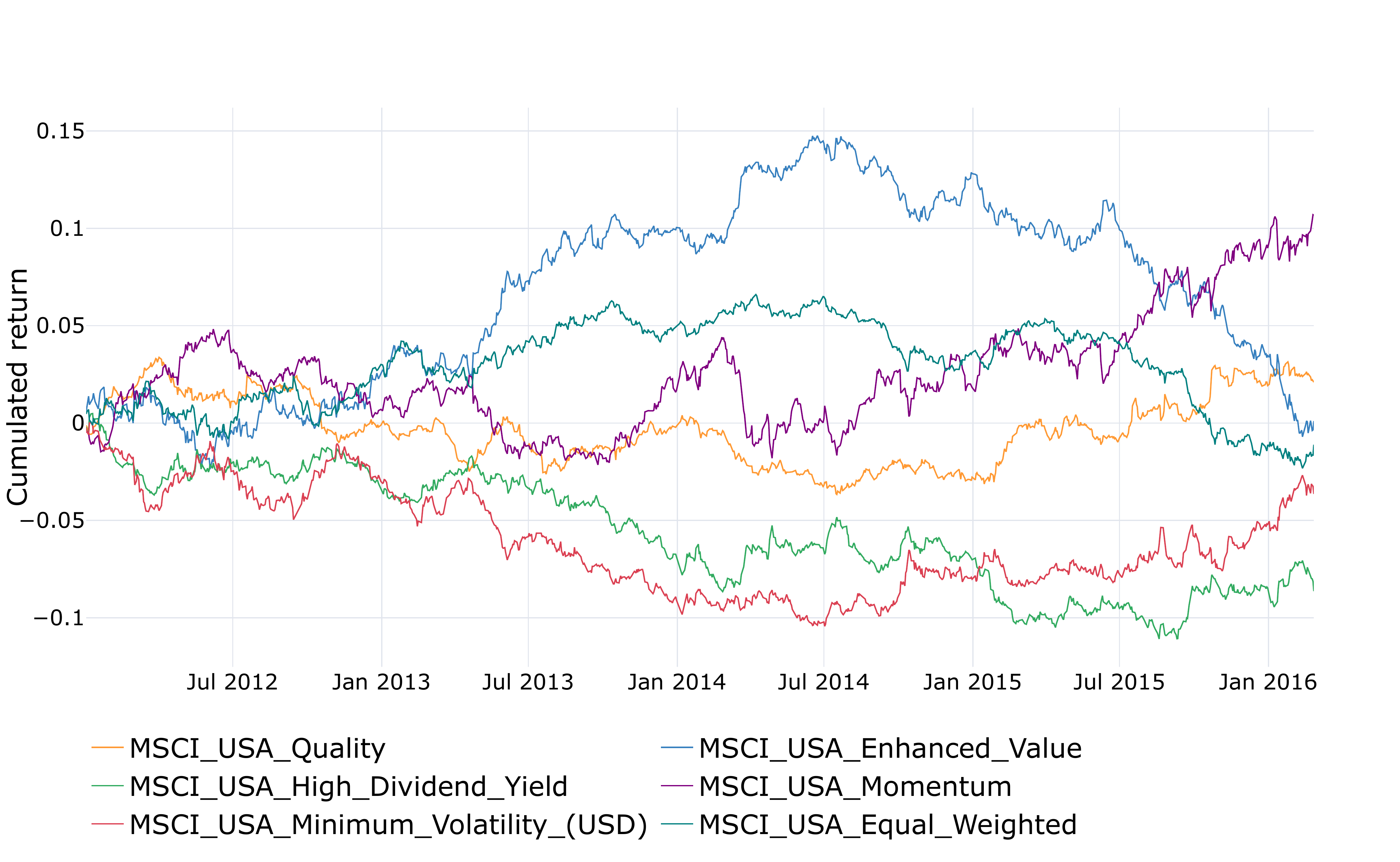}
  \caption{Cumulative returns of MSCI USA enhanced factors. Returns are in excess of the market in USD, for the date range Jan 2012 to Feb 2016.}
  \label{fig:MSCI-index}
\end{figure}

\begin{table}
	\caption{Description of datasets.}
	\ra{1.2}
	\centering
	\footnotesize
	\begin{tabular}{@{}cccc@{}}\toprule
		Dataset & Date & Nr of features & Frequency\\
		\cmidrule{1-4} 
		Factor data & Jan-1988 to Feb-2016 & 25 & Daily\\
		MSCI Enhanced & Jan-1999 to Feb-2016 & 6 & Daily\\
		\bottomrule
	\end{tabular}
	\label{table:datasets}
\end{table} 

\section{Dynamic asset allocation system}
\label{section:DAA_system}

Investment on single factor strategies has been shown to have significant returns over the long term but how to build multi-factor strategies and rotate factors according to market conditions is not straightforward. Factor indices are time series data, hence we take advantage of the capacity of hidden Markov models to identify underlying regimes in sequences of observations and build a dynamic asset allocation system. We will first determine the optimal number of hidden states to model market regimes and then, in order to avoid excessive transactions costs through frequent rebalancing, we optimize the rebalancing signal. 

\subsection{DAA system}
\begin{figure}[h]
\centering
  \includegraphics[width=0.5\textwidth]{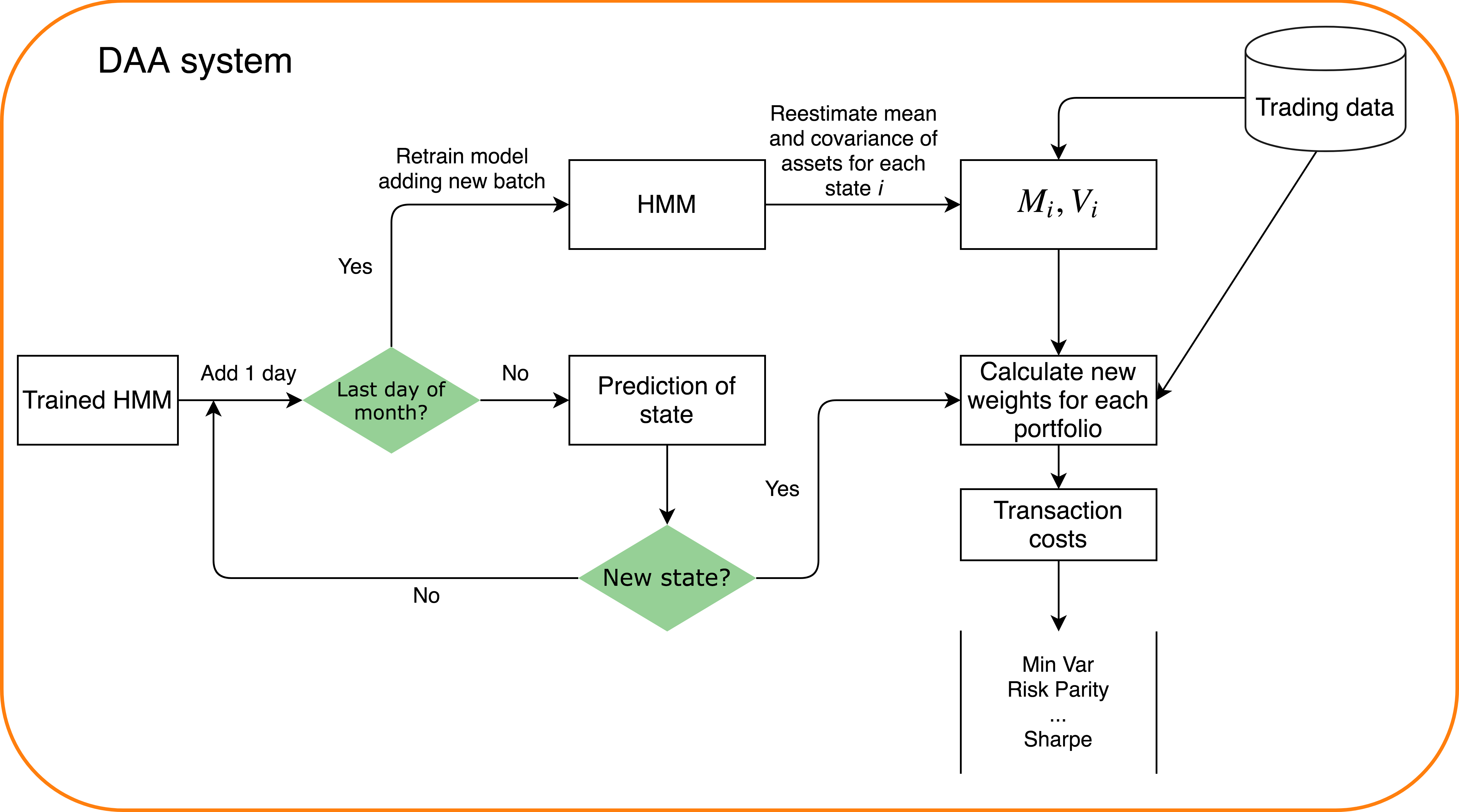}
  \caption{Dynamic Asset Allocation system diagram.}
  \label{fig:DAAsystem}
\end{figure}

\begin{figure}[h]
\centering
  \includegraphics[width=0.3\textwidth]{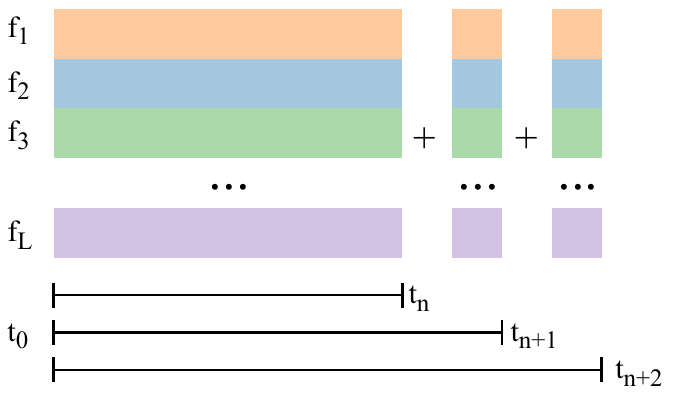}
  \caption{Data scheme.}
  \label{fig:expWindow}
\end{figure}

We design a dynamic trading framework with daily evaluations and monthly re-adjustments as shown in figure \ref{fig:DAAsystem}. Each day a new vector of returns is added to the training set with an expanding window, and the state is predicted. Returns are lagged by one day in order to avoid look-ahead bias. 
Because this prediction is noisy, we'll determine an optimal window of consecutive days in the new state before the portfolio is rebalanced. Once a change of state has been accepted, the vector of means and covariance matrix from the new state are retrieved and the portfolio weights optimized, with transaction costs calculated after the rebalance. 
After a full month has passed, we add this new batch of data to the training set with an expanding window and retrain the model. Figure \ref{fig:expWindow} shows how data is added daily with an expanding window.
While this will not produce immediate changes in the model parameters (transition matrix and emission distributions) in time they should change slightly to accommodate the new information. Therefore, we can capture changes on the dynamics of the system over time.

\subsubsection{Model selection}

The number of latent states in a HMM has to be set in advance, before training. One option is to use the Bayesian Information criterion (BIC), a penalized log-likelihood function that can be used for model selection \citep{schwarz1978}. BIC is defined by:
\begin{equation*}
	BIC = -2 \log p(D|\hat{\theta}) + d \log(N)
\end{equation*}
where $d$ is the number of free parameters in the model and $N$ is the number of samples. Thus, calculating the score over a range of $K$ states, we can select the model with the lowest value. Another option is to follow a greedy approach, calculating performance of the portfolios built with a different number of regimes and selecting the model with highest performance.   

\begin{figure}[h!]
  \includegraphics[width=0.5\textwidth]{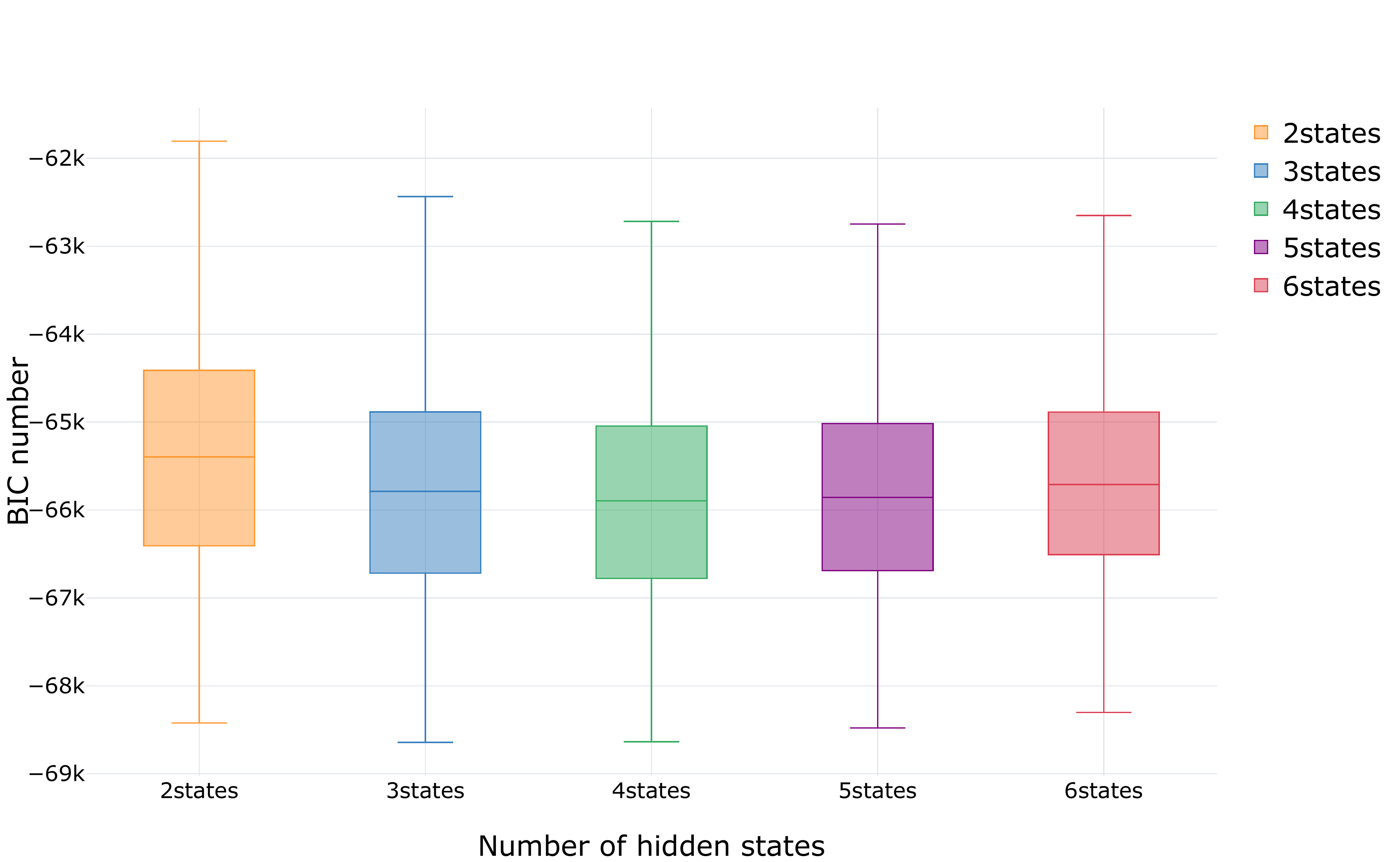}
  \includegraphics[width=0.5\textwidth]{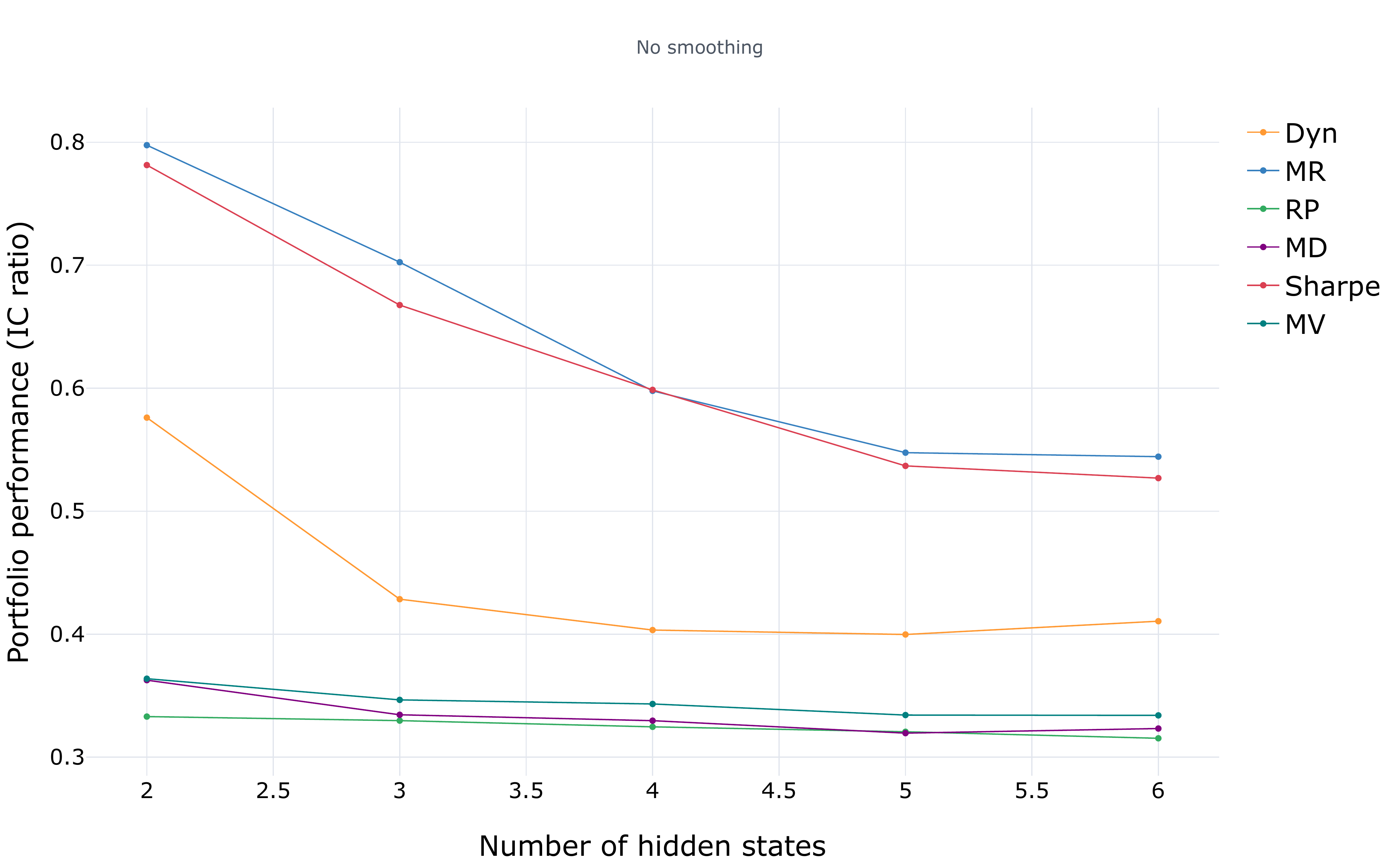}
  \caption{The Top plot shows the boxplot of BIC number for different number of states: a two state model has a higher BIC but there is no distinction between three, four and five; the Bottom plot shows performance of portfolios as a function of number of hidden states.
  the two state model yields a better performance for the majority of portfolios.}
  \label{ICvsSTATES}
\end{figure}

In the financial HMM literature \citep{Guidolin2008}, regime switching models normally range between two and four states. Keeping the number of states low allows better interpretability, so we selected 200 random combinations of 5 assets each and used this combinations to train an HMM with 2, 3, 4, 5 and 6 hidden states respectively. From each HMM information we built different types of portfolios, as will be explained in section \ref{subsection:tradeStrategies}. The performance of each portfolio was calculated using the IR ratio (the ratio between annualized return and annualized volatility); the plots of BIC and performance as a function of number of states are shown in Figure \ref{ICvsSTATES}. The BIC score is quite similar for states three to six (four being the lowest) and is slightly higher for two states. While this would suggest use of a four regime model, performance of portfolios for three and four states is significantly lower than for two states, so we have selected a two-state model. Two-state models can be interpreted as expansion-contraction.

\subsubsection{System calibration}

The dynamic asset allocation system requires a trained HMM to model regime changes and the selection of an optimal time window to decide when a change of state has taken place and the portfolio has to be rebalanced. 

For the first part of the work, where we want to test if the proposed DAA system adds value to multi-factor strategies, we test it using multiple combinations of factors, and calibrate the system for each combination. 
From a pool of 25 factor indices we select $n$ assets randomly and use their returns to train a HMM. As factors can be grouped into five families (following table \ref{table:JPM}), we randomly select one factor from each group so all families are represented. This yields a total of 1260 combinations. We then use the same factors to build the portfolios.

\begin{figure}[ht!]
\centering
  \includegraphics[width=0.5\textwidth]{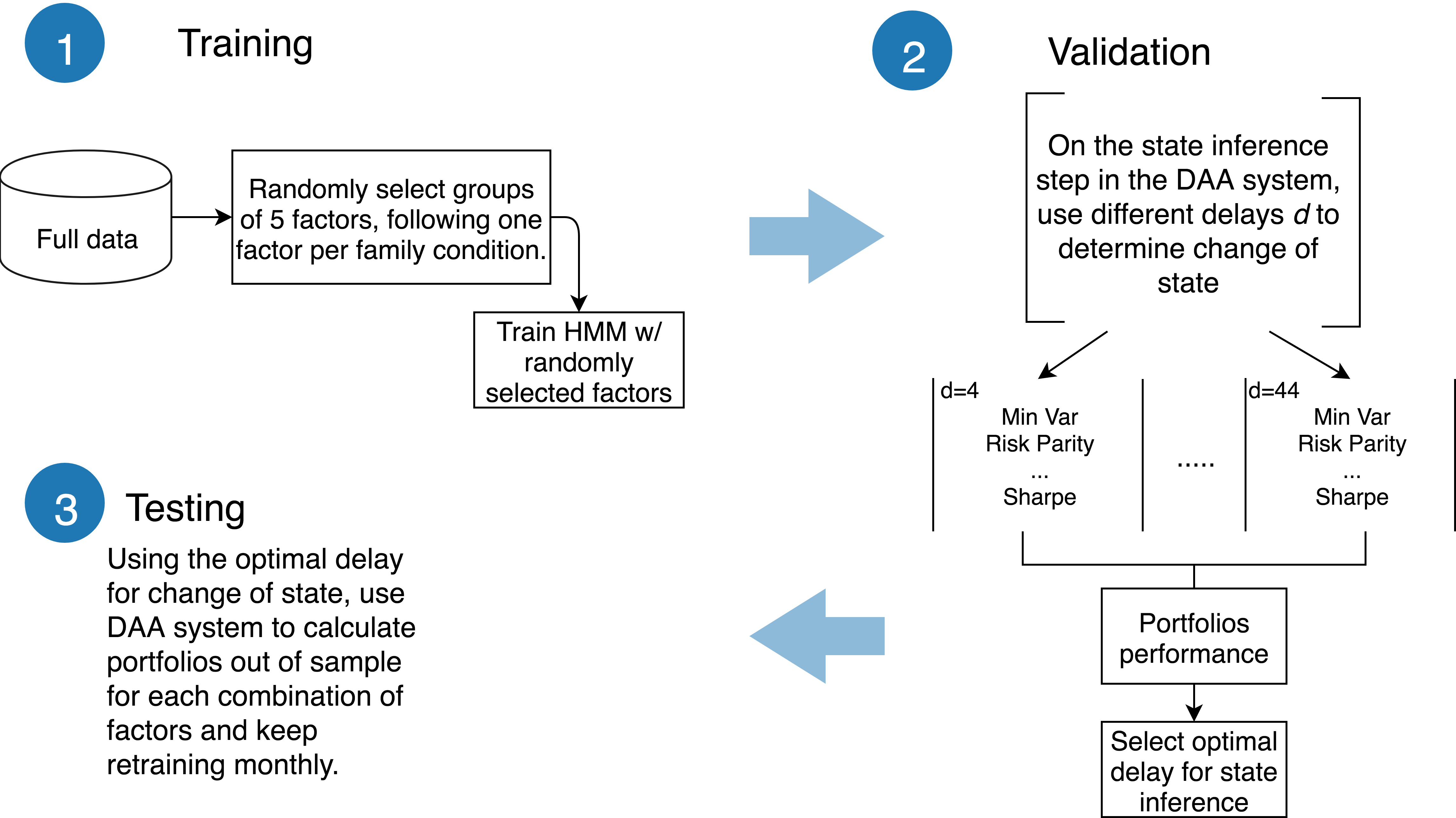}
  \caption{Full schematic of calibration and usage of the dynamic asset allocation system for smart beta investing.}
  \label{fig:hmm-flowchart}
\end{figure}

We divide the data set into three parts, training (15 years), validation (9 years) and test set (4 years). In order to avoid getting stuck in a local maximum we do random initialization with initial parameters calculated from the training data and select the model with highest score. Figure \ref{fig:hmm-flowchart} shows the process of training, validation and test using the DAA system.

\begin{figure}[ht!]
\centering
  \includegraphics[width=0.5\textwidth]{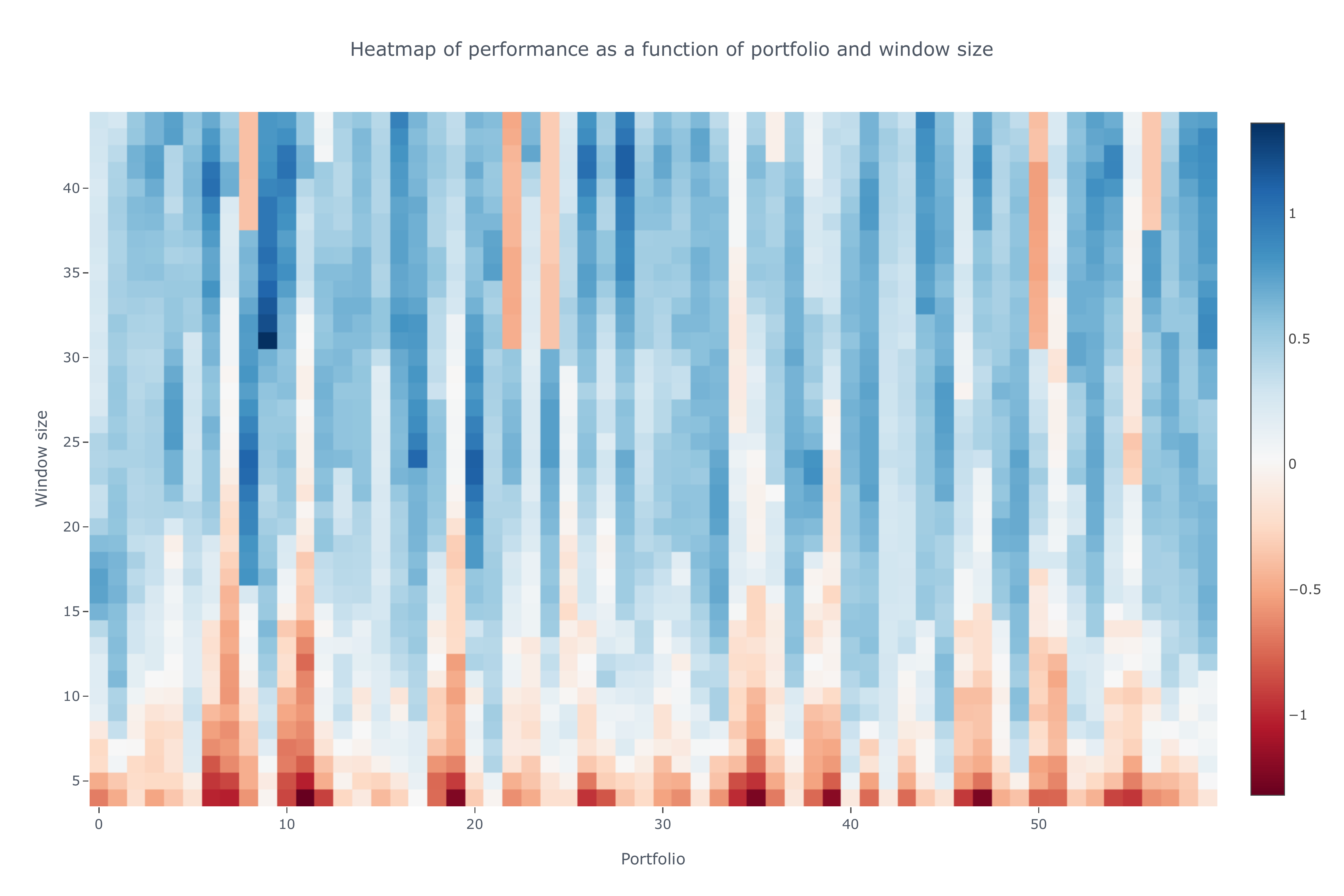}
  \caption{A subset of the 1260 portfolios is plotted. The colormap corresponds to the performance measured by IR (adjusted for transaction costs) as a function of window size.
 In the majority of cases performance is low for smaller windows due to frequent re-balance; performance tends to improve with window size, 15. However, if the window is too large, performance may decrease again as it fails to take advantage of more frequent regime changes.}
  \label{fig:heatmap}
\end{figure}

The regime prediction is done by passing the whole series of returns up to the previous day to decode the most probable sequence of hidden states, and keep the last value as the state prediction. This daily prediction is noisier that it would be if a whole month of returns was passed together, and we cannot re-balance a portfolio each time a change of state is flagged, as quite often this would mean a daily re-balance. Instead, in the validation set, we look for a window of $d$ consecutive days in the same new state and then we flag a change of regime and re-balance the portfolio accordingly. Figure \ref{fig:heatmap} shows the performance of a selection of portfolios as a function of the time window $d$. While certain combinations of assets perform consistently better than others with larger windows, smaller windows have the worst performance in all cases. The main reason is that performance of portfolios is adjusted for transaction costs, so smaller windows mean higher portfolio turnover and therefore, higher costs. We use the training set to identify the optimal window for each combination of assets. 

\subsection{DAA system with Feature Saliency: FS-DAA}
\label{sec:feature-selection}

So far, we proposed a DAA system where the time series to train the HMM were known in advance, which can be a limitation. Therefore, we propose a novel DAA system that incorporates an embedded feature selection method during the training, by using a Feature Salience Hidden Markov Model (FSHMM) as described in section \ref{section:FSHMMtheory}. This method allows to select features that contribute to the regime identification, called regime dependent, and rejects features that don't depend on the regimes.

\begin{figure}[ht!]
\centering
  \includegraphics[width=0.5\textwidth]{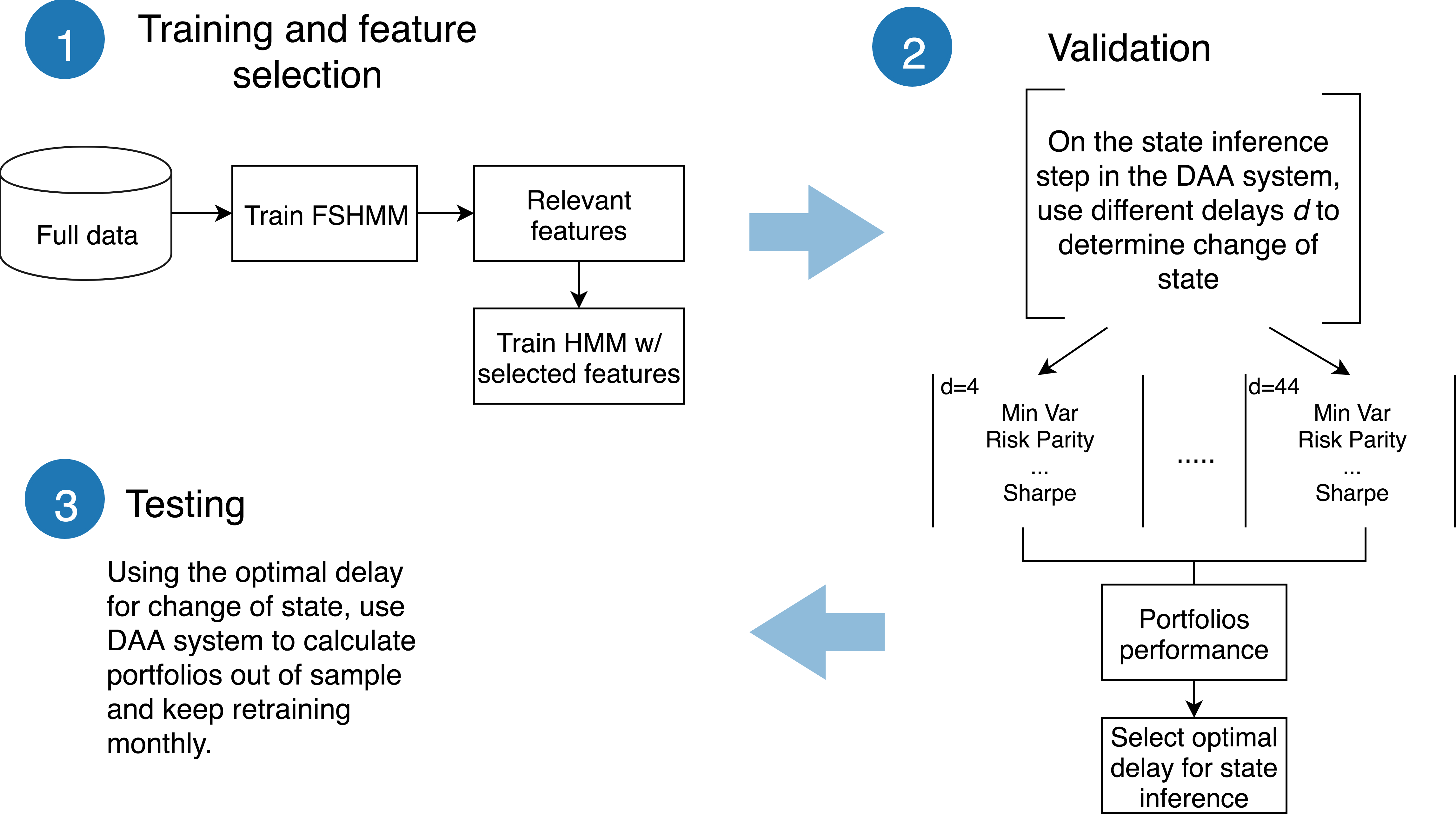}
  \caption{Full schematic of calibration and usage of the DAA system with embedded feature selection for smart beta investing.}
  \label{fig:fshmm-flowchart}
\end{figure}

Figure \ref{fig:fshmm-flowchart} shows the different stages for training, validation and test using this new DAA system, that we called FS-DAA.
FS-DAA takes multiple time series data and fits a FSHMM, that assigns a saliency to each time series. Higher saliency means that the feature is selected. Because FSHMM proposes that features are conditionally independent, the fitted model has diagonal covariance matrices. We therefore take the selected relevant features and used them to train a HMM with full covariance matrices. 

As a first step to assess whether FSHMM can distinguish between relevant features and noise, we generated irrelevant features of random noise and added them to our daily factor data set. We tested this using different number of features, number of observations and values of $k_l$. For each case, $k_l$ was the same for all features, both relevant and noise. Results are summarized in Tables \ref{table:rhos1} and \ref{table:rhos2}. In all cases, the algorithm assigned low values of saliency for the irrelevant features and high values for the relevant ones. 

Secondly, we train a DAA system using all 25 features from the factor dataset, and we train a FS-DAA system that takes the 25 features, selects the relevant ones and then trains a HMM only with those factors and compare the regimes obtained. Finally, using these two systems, we build a strategy using a MSCI USA enhanced family of factor indices. Both models are trained using 16 years of data (from 1990 to 2006) and then retrained every month until 2016. We use 7.5 years of trading data to estimate mean and covariance of the MSCI indices for each regime, from Jan 1999 to June 2006, to have a robust estimation of the covariance matrix for both regimes. We then use a validation set of 6 years to select the optimal time window to set a change of state, and a test set of 4 years. 

One advantage of the proposed DAA system is that it allows to decouple data used to train the HMM to detect regimes from the data used for allocation. This is useful for factor investing because we can build factors with a long history (as the factor dataset) and then use real life, investable assets that have a shorter history (MSCI enhanced data) to build the portfolios.

\section{Results and analysis}
\label{section:results}

Firstly, the DAA system performance is compared with baseline strategies on the large factor dataset. Then, the implementation of FSHMM algorithm is discussed. Lastly, we test the proposed FS-DAA system with real life assets using the MSCI indices dataset.

\subsection{Trading strategies and benchmarks}
\label{subsection:tradeStrategies}

Instead of constructing only one kind of portfolio we build several: Risk Parity, Maximum diversification, Minimum Variance, Max return, Max Sharpe and a modified max return - (for a short description of each portfolio, see Appendix (\ref{appendix:port}). Risk Parity (RP), Maximum diversification (MD) and Minimum Variance (MV) are constructed taking into account only the covariance matrix, so they can be considered more risk aware. Max return (MR), Max Sharpe (Sharpe) and modified max return (Dyn) all consider the mean of the return during the construction, so they tend to be more aggressive. 

For comparison we built an equally weighted portfolio and a benchmark for each asset combination.
Each benchmark is constructed using the same optimization method as its DAA system counterpart, but are rebalanced monthly and the covariance matrix is estimated using ``single regime'' past returns. The DAA-system instead has two covariance matrices, one for each regime. 
All portfolios and their benchmarks are constructed taking into account transaction costs. Costs are calculated by multiplying portfolio turnover (how much a portfolio is rebalanced) with a transaction cost of 50bps (0.5$\%$), for each selling and buying.

\subsection{DAA system compared to baseline}

We first evaluated our DAA system by using 1260 combinations of randomly selected assets to train the HMM and for the allocation, and compare it with their benchmarks.

\begin{figure}[h!]
\centering
  \includegraphics[width=0.5\textwidth]{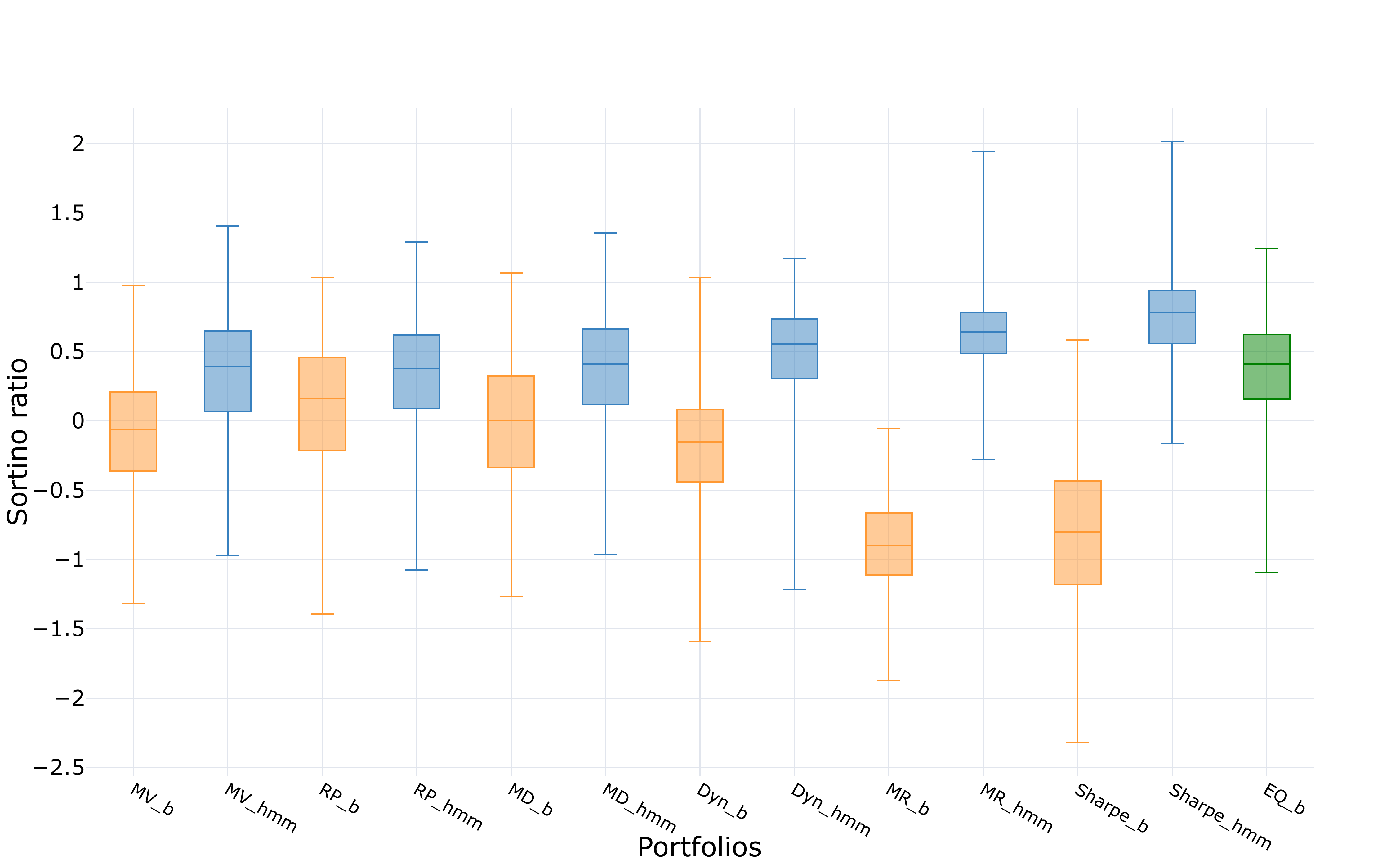}
  \caption{Boxplots corresponding to the Sortino ratio for all portfolios calculated using a HMM (blue) and their benchmarks (orange) and an equally weighted portfolio (green).}
  \label{fig:boxplots_all_port}
\end{figure}

Figure \ref{fig:boxplots_all_port} shows the performance measured through Sortino ratio of all portfolios calculated using the DAA system, and their benchmarks. We can see that all portfolios constructed using regime information perform better than their counterpart. Portfolios that are more return-oriented because are calculated using the mean returns in the optimization process improve greatly with respect to their benchmarks while more risk focused portfolios show an improvement with respect to their single-regime counterparts but show a similar performance to equally weighted portfolios. 

\begin{figure}[h!]
  \includegraphics[width=0.5\textwidth]{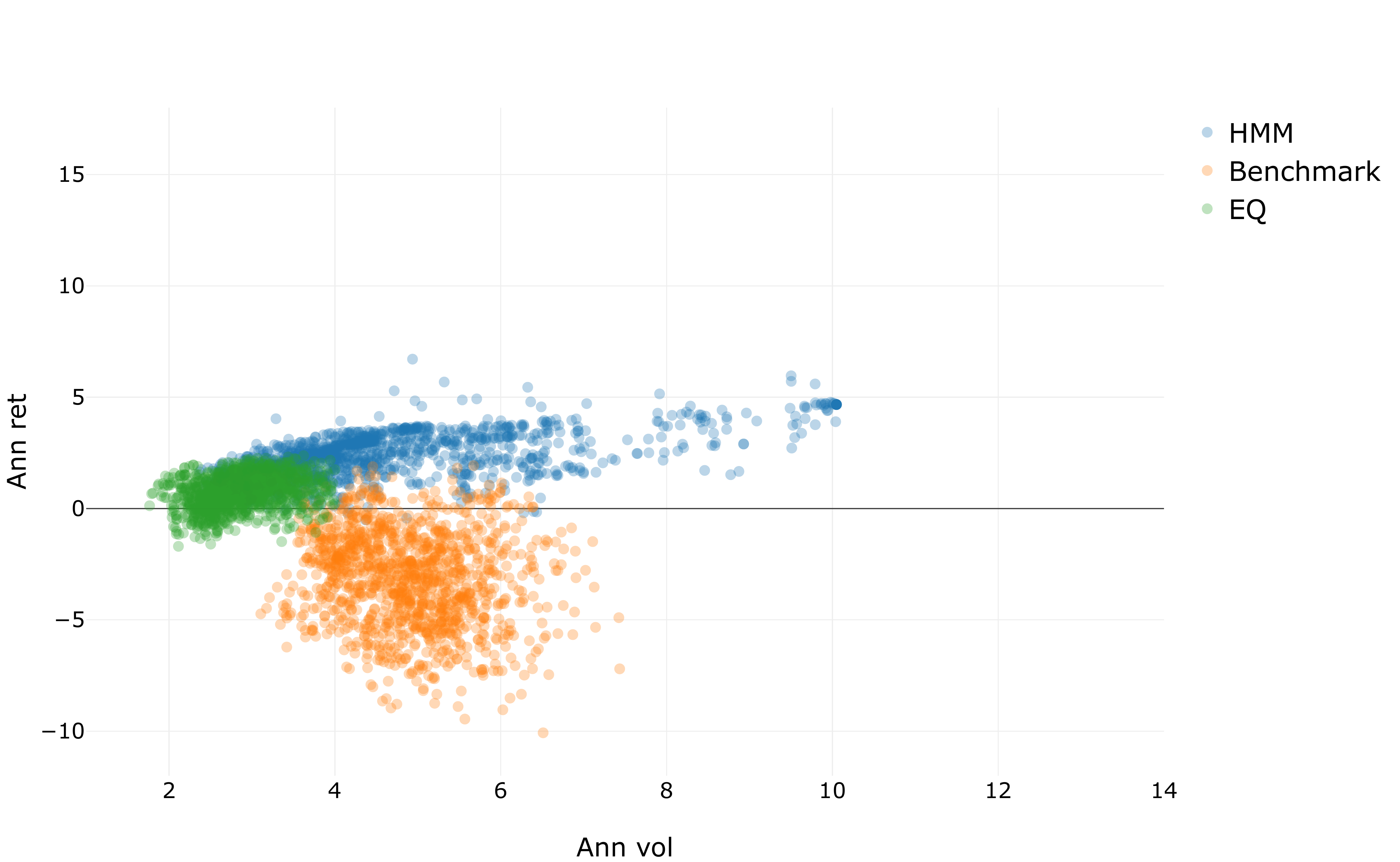} \\
  \includegraphics[width=0.5\textwidth]{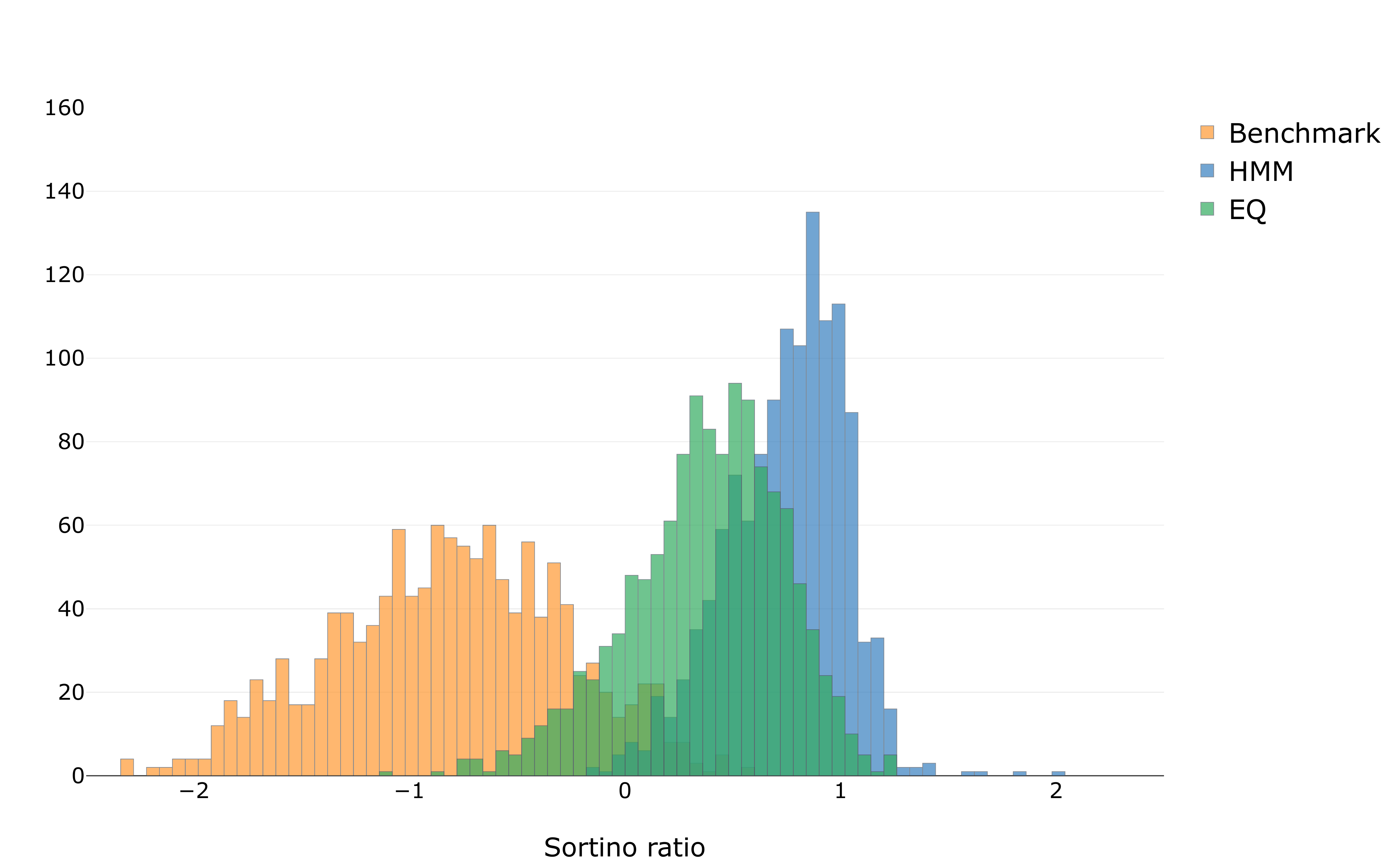}
  \caption{Left plot shows annualized return as a function of annualized volatility for Sharpe portfolios built using HMM information (blue), Sharpe portfolios rebalanced monthly (orange) and EQ portfolios (green). Right plot corresponds to the Sortino distribution of the plots. All plots correspond to the test set (are out of sample).}
  \label{fig:RV}
\end{figure}

The highest performing portfolio is Sharpe, that takes into account both mean and covariance in the construction process. Figure \ref{fig:RV}-Top  shows the annualized return as a function of annualized volatility for the Sharpe portfolios and their benchmarks. Portfolios built using HMMs show a higher return and less volatility than their unconditional counterpart, and higher return and volatility than the EQ portfolios. 
Figure \ref{fig:RV}-Bottom shows a risk adjusted return metric (Sortino) for the same portfolios. We can see that the HMM portfolios yield a better performance than their benchmarks. 

\rowcolors{2}{gray!8}{white}

\begin{table*}
\caption{Average performance of portfolios built using HMMs and their benchmarks. Top portfolios that are more aggressive have a higher risk adjusted return (measured through IC and Sortino ratios) than their unconditional counterpart and the equally weighted portfolio. Bottom portfolios that are more defensive (only the covariance matrix is taken into account in the construction process) perform worse than their benchmark counterparts and the EQ portfolio.}
	\ra{1.2}
	\centering
	\small
	\begin{tabular}{@{}rlccccccccc@{}}\toprule
							& Ann ret & Ann vol & IR & Skw & kurt & D. risk & Sortino & DD & DD days \\
		\cmidrule{1-1} \cmidrule{2-10}
  		EQ					& 0.77   & 2.88  & 0.26  & -0.14  & 0.81  & 2.05  & 0.37   & 379   & 318 \\      
        Dyn HMM	  			& 1.67   & 4.73  & 0.34  & -0.19  & 1.35  & 3.37  & 0.48   & 32    & 291 \\
		Dyn Bench 			& -0.60  & 3.98  & -0.14 & -0.40  & 1.68  & 2.96  & -0.19  & 1136  & 682 \\
		Sharpe HMM			& 2.31   & 4.66  & 0.53  & -0.19  & 1.16  & 3.29  & 0.75   & 429   & 253 \\
		Sharpe Bench		& -3.14  & 4.89  & -0.64 & -0.79  & 4.49  & 3.80  & -0.82  & 1375  & 873 \\
		MR HMM 				& 3.190  & 7.03  & 0.46  & -0.19  & 1.34  & 4.98  & 0.65   & 35    & 264 \\
		MR Bench			& -5.03  & 7.20  & -0.69 & -0.78  & 3.71  & 5.63  & -0.88  & $>$4000 & 1001 \\
		MV HMM				& 0.61   & 2.41  & 0.24  & -0.14  & 0.96  & 1.72  & 0.35   & 662   & 309 \\
		MV Bench			& -0.12  & 2.24  & -0.07 & -0.11  & 0.83  & 1.61  & -0.09  & 520   & 511 \\
		MD HMM				& 0.69   & 2.54  & 0.26  & -0.14  & 1.01  & 1.80  & 0.37   & 340   & 306 \\
		MD Bench			& 0.01   & 2.39  & -0.02 & -0.12  & 0.84  & 1.71  & -0.02  & 454   & 447 \\
		RP HMM				& 0.63   & 2.58  & 0.24  & -0.13  & 1.04  & 1.84  & 0.34   & 212   & 302 \\
		RP Bench			& 0.20   & 2.40  & 0.07  & -0.13  & 1.04  & 1.72  & 0.10   & 475   & 416 \\
		\bottomrule
	\end{tabular}
\label{table:results-part1}
\end{table*}

Table \ref{table:results-part1} shows different performance metrics averaged for each type of portfolio. In most cases, HMM-portfolios show better performance than their unconditional benchmarks on all metrics, and more return-oriented portfolios perform better than equally weighted ones. Performance improvement comes both from higher returns and risk reduction in return-oriented portfolios. Additionally, skewness and kurtosis are lower than benchmark returns and maximum drawdown is lower (and for a shorter period of time) in most cases.

\subsection{DAA system with FSHMM}

\begin{figure}[h!]
  \includegraphics[width=0.5\textwidth]{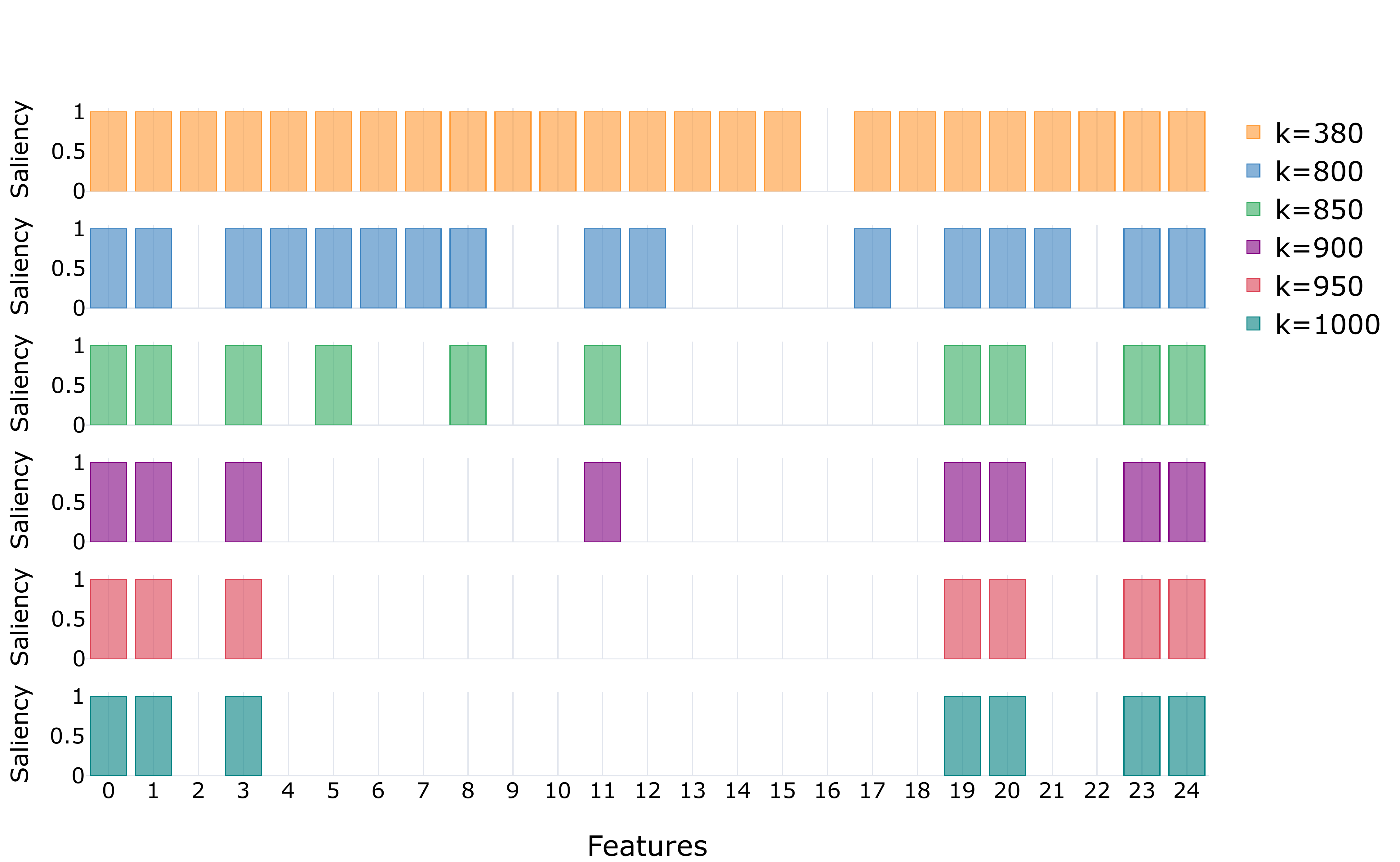}
  \caption{Selected features in the training set ($T = 3800$ observations) of the 25 factor return series with different values of $k$. With small values of $k$ all features are accepted. With $k\geq T/4$ the algorithm selects a relevant subset of features.}
  \label{fig:rhos}
\end{figure}

We then used the algorithm to detect relevant features in our data set of 25 factor indices. Figure \ref{fig:rhos} shows the feature saliencies of all factor return series for different values of $k$. As the training set has about 3800 observations, we chose values of $k$ closer to a quarter of that number following the heuristics proposed in \citet{FSHMM:article}. The selected features are: Book Value Yield, 1 Yr Fwd Earnings Yield, Sales Yield, 6 Month Price Momentum, 12 Month Price Momentum, EPSCV, Beta. This is of interest as the selected factors represent four of the six or seven factor families mentioned in section \ref{subsection:smartBeta}.

\begin{figure}[h!]
  \includegraphics[width=0.5\textwidth]{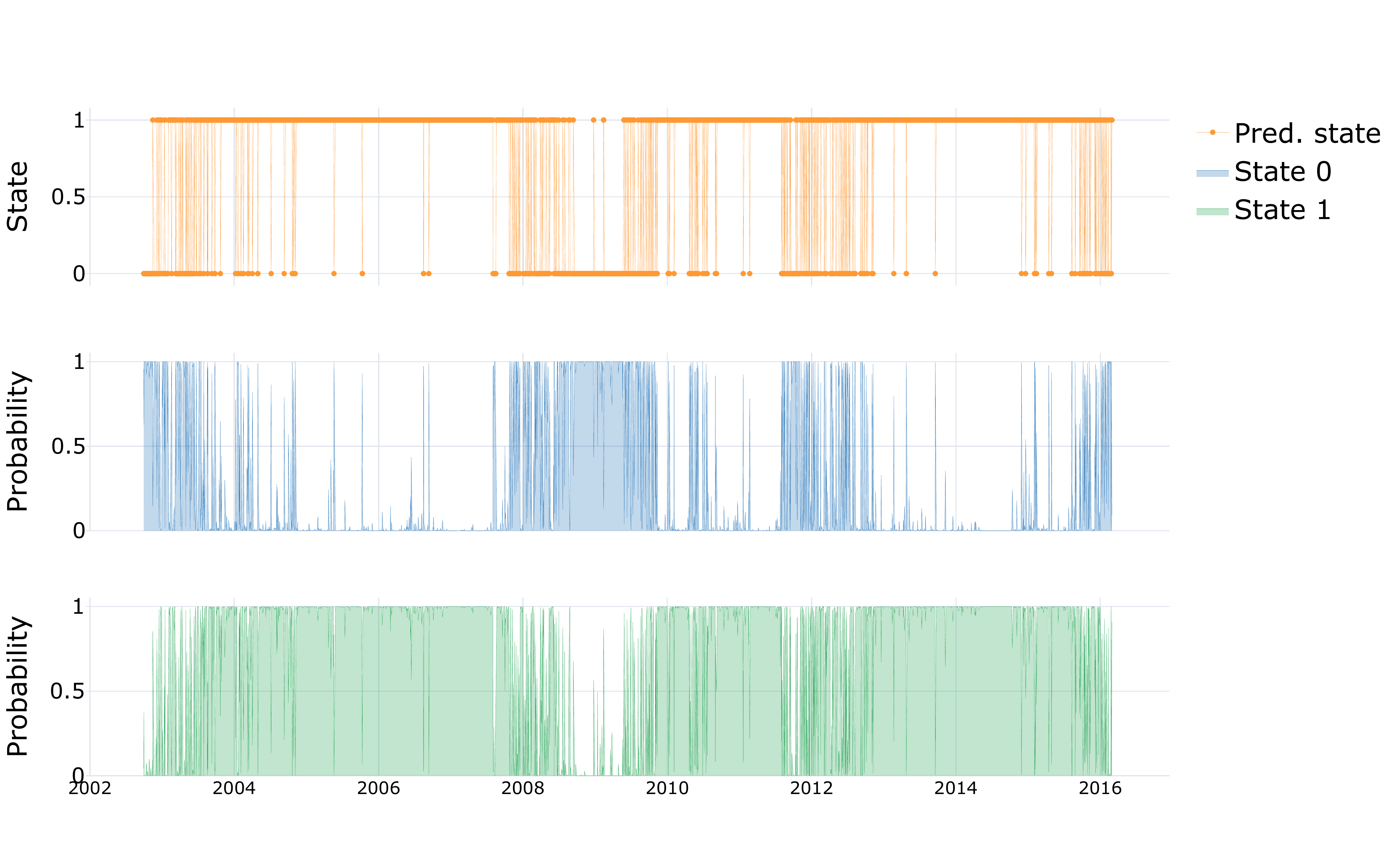}\\
  \includegraphics[width=0.5\textwidth]{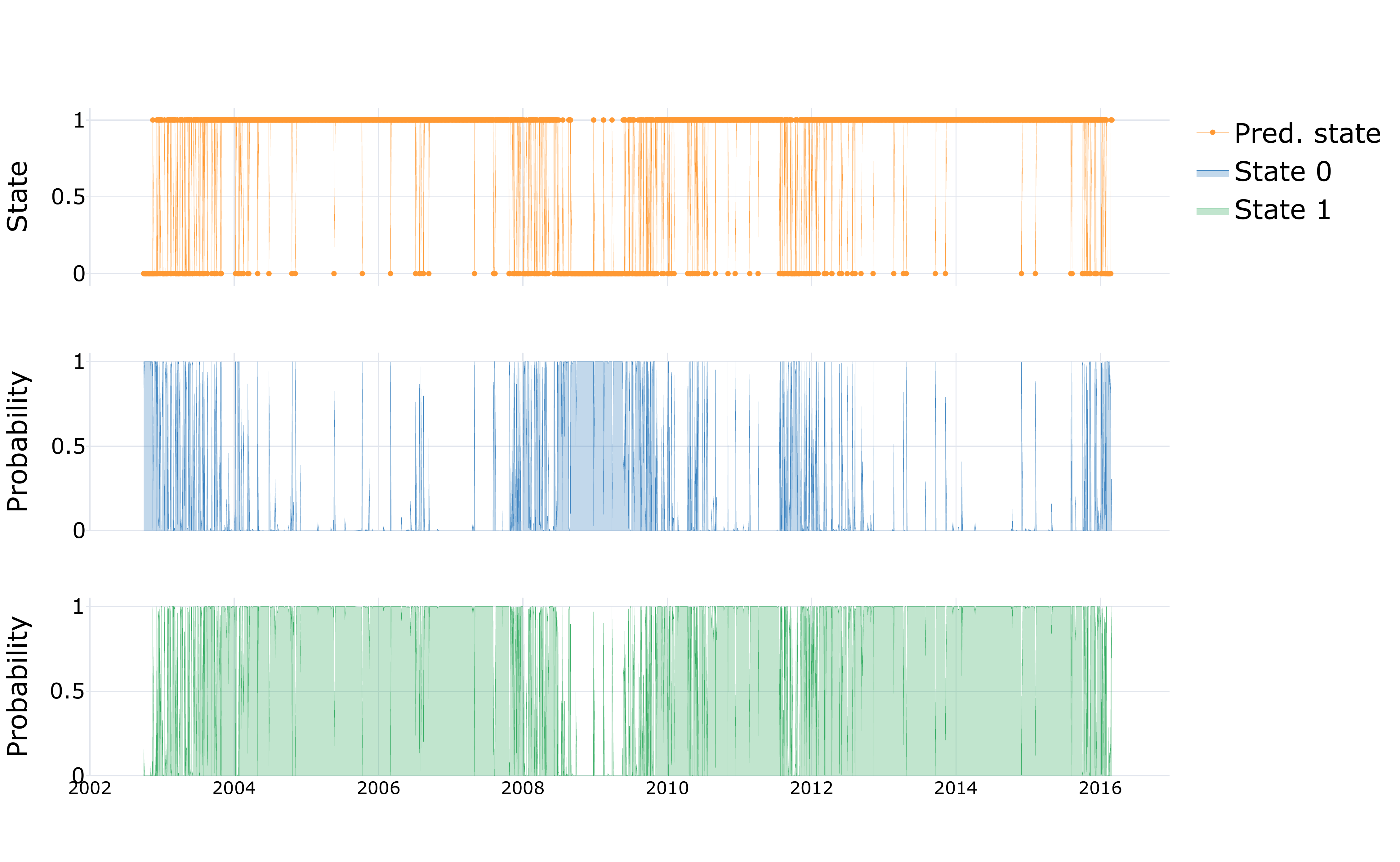}
  \caption{Left Plot corresponds to predicted state and state probabilities for the model trained with relevant features. Right Plot corresponds to the HMM trained with all 25 features.}
  \label{fig:filtered-prob}
\end{figure}

For comparison, we trained a HMM using all 25 feature and a model trained with the selected assets. Figure \ref{fig:filtered-prob} shows the predicted state and estimated probabilities for the model after training. We can identify state 1 as a "good state", and state 0 as a "bad" state. The plots clearly identify the 2008 economic crisis - the first steps developed in August and September of 2007 with some episodes between January and May 2008 before the big crash in September 2008. Both models identify spikes of state 0 in the second half of 2007 and transition fully to state zero during 2008. The model trained with relevant features tends to be more sensible to the distress state - it spends 24$\%$ of the time in this state versus $20\%$ of the model trained with the full set of features. The average duration of state 0 is 3.8 days vs average length of 3.2 days of the full model. No smoothing was applied to the predicted probabilities to calculate these values. 

\subsection{DAA-FS system with MSCI indices}
\label{section:MSCI-strategy}

In this section we evaluate performance of the DAA-FS system using a subset of factors from the daily factor dataset after feature selection, and MSCI enhanced factors for allocation, and compare it with the DAA system without feature selection, that trains the HMM with all 25 factors from the dataset. 

\begin{figure}[h!]
  \includegraphics[width=0.5\textwidth]{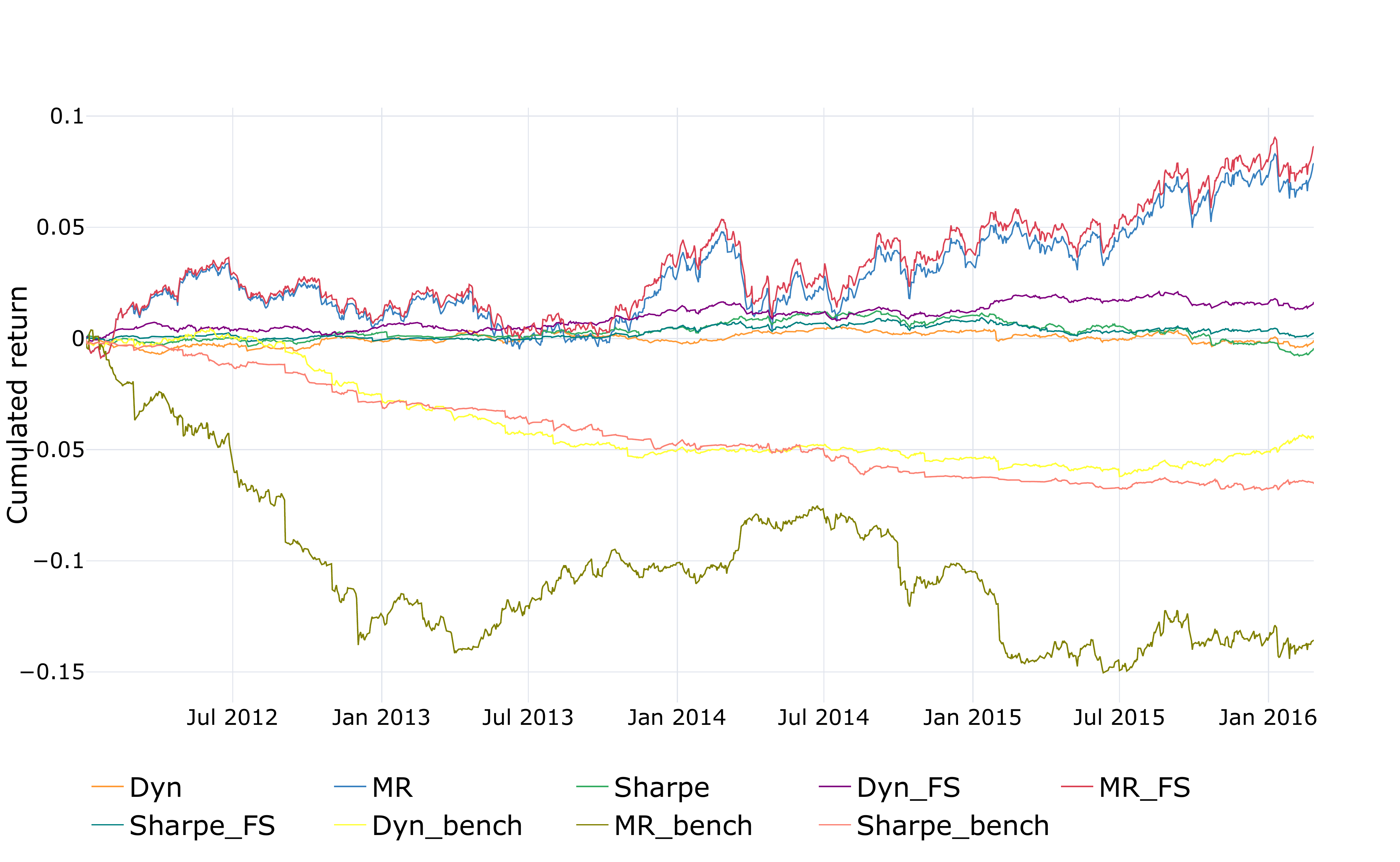}\\
  \includegraphics[width=0.5\textwidth]{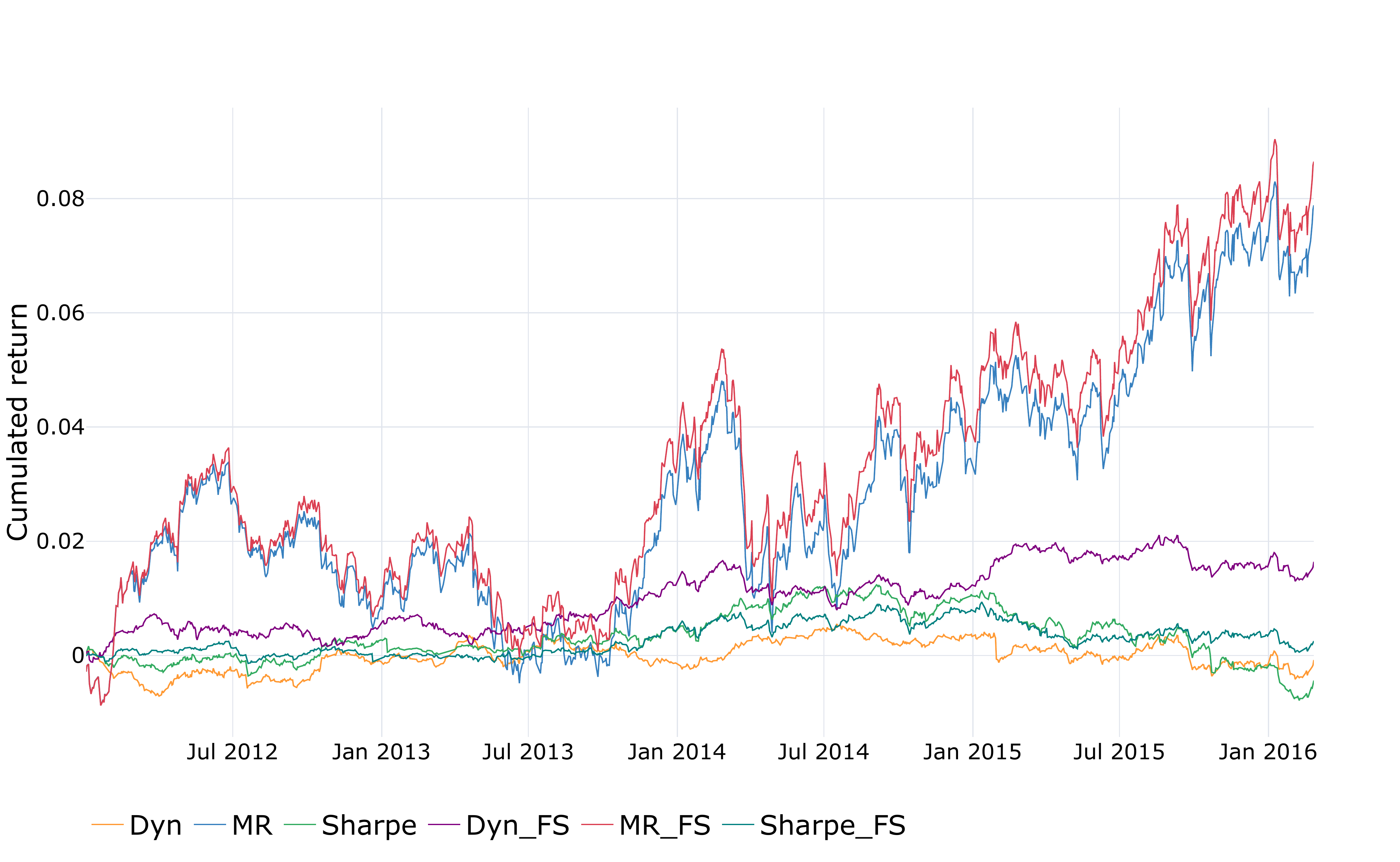}
  \caption{Top plot corresponds to portfolios built using information from HMM with feature saliency, portfolios built using information from HMM with full features and their benchmarks. Both HMM portfolios accumulate higher returns than the benchmarks. Bottom plot shows that cumulative returns of FSHMM and fullHMM, portfolios built using FS have a better performance. Returns are in excess of the market in USD, for the period Jan 2012 to Feb 2016.}
  \label{fig:MSCI-HMM-port}
\end{figure}

For simplicity we calculated only Sharpe, MR and Dyn portfolios, as they showed a significantly better performance when using a regime switching model in their construction than risk-focused portfolios and their benchmarks. Figure \ref{fig:MSCI-HMM-port} shows the cumulative return of these three portfolios with a full feature HMM, FSHMM and the benchmarks constructed without regime information. Both HMM portfolios perform better than their benchmarks (top plot) and portfolios constructed using an HMM with feature selection perform slightly better than portfolios built with a full feature HMM (bottom plot). 

Metrics performance for all portfolios and for the MSCI enhanced indices net of market are shown in table \ref{table:MSCI-metrics}. All metrics are annualized and are out-of-sample, covering the period Jan-2012 to Feb-2016. The results obtained using DAA and FS-DAA show a robust improvement with respect to their benchmarks. We can see that only three MSCI indices have a positive IR in the period, and two of the three FSHMM portfolios show the highest IR in all cases. Reduction of downside risk is achieved in most cases that use either a full-feature HMM or a FSHMM with respect to their benchmarks and the MSCI indices.

\begin{table*}
	\caption{Metrics for portfolios built using FSHMM, all assets (HMM), their benchmark and the individual MSCI indices used to build the portfolios. The metrics covered the period Jan 2012 to Feb 2016. }
	\ra{1.2}
	\centering
	\small
	\begin{tabular}{@{}rlccccccccc@{}}\toprule
									& Ann ret & Ann vol & IR & Skw & kurt & D. risk & Sortino & DD & DD days \\
		\cmidrule{1-1} \cmidrule{2-10}
		Sharpe FSHMM				& 0.061 & 0.50  & 0.12  &-0.71  & 2.85  & 0.37  & 0.16  &-94  & 387 \\
		Sharpe HMM					&-0.11  & 0.65  &-0.16  &-0.70  & 3.84  & 0.49  &-0.22  &-164 & 522 \\
		Sharpe Bench				&-1.62  & 0.92  &-1.76  &-2.75  & 15.0  & 0.82  &-1.98  &19825& 1452\\
        Dyn FSHMM	  				& 0.39  & 0.65  & 0.61  &-0.41  & 0.84  & 0.47  & 0.84  &-52  & 141 \\
		Dyn HMM						&-0.019 & 0.60  &-0.032 &-1.12  & 9.03  & 0.45  &-0.042 &-175 & 566 \\
		Dyn Bench					&-1.10  & 1.03  &-1.07  &-2.76  & 16.2  & 0.88  &-1.24  &-1508& 1123\\
		MR FSHMM 					& 2.02  & 3.20  & 0.63  &-0.39  & 1.83  & 2.30  & 0.88  &-82  & 62  \\
		MR HMM 						& 1.85  & 3.19  & 0.58  &-0.39  & 1.84  & 2.29  & 0.80  &-92  & 62  \\
		MR Bench					&-3.46  & 3.78  &-0.91  &-2.71  & 20.5  & 3.17  &-1.09  &-4032& 1250\\
		MSCI Quality				& 0.50  & 2.76  & 0.18  & 0.20  & 2.02  & 1.90  & 0.26  &-208 & 837 \\
		MSCI Enhanced Value			& 0.025 & 3.97  & 0.0064& 0.029 & 0.86  & 2.83  & 0.0090&-105 & 599 \\
		MSCI High Dividend Yield	&-2.16  & 3.22  &-0.67  & 0.38  & 0.85  & 2.24  &-0.96  &-2374& 1317\\
		MSCI Momentum				& 2.48  & 4.35  & 0.57  &-0.35  & 1.42  & 3.11  & 0.80  &-144 & 475 \\
		MSCI Minimum Volatility		&-0.89  & 3.58  &-0.25  & 0.10  & 0.69  & 2.52  &-0.35  &-38371& 906\\
		MSCI Equal Weighted			&-0.27  & 2.94  &-0.092 &-0.045 & 0.74  & 2.09  &-0.13  &-135 & 675 \\
		\bottomrule
	\end{tabular}
	\label{table:MSCI-metrics}
\end{table*}

\section{Conclusions and future work}
\label{section:conclusion}

The main focus of the paper is to improve smart beta strategies through the use of regime switching models. The main contributions from this work are:

\begin{enumerate}
\item We have shown that constructing a portfolio using information from a HMM with two latent states trained with the same assets that will be used for allocation, improves performance with respect to the same portfolio built with a single regime approach. 

We have tested this by calculating different types of portfolios, ranging from more risk focused to more aggressive. 
The improvement is more significant for return-oriented and balanced portfolios where return or risk-adjusted return is optimized achieving on average an information ratio of 50$\%$ annually in excess of market, and is less evident in risk-focused portfolios (Risk Parity, Minimum Variance and Maximum diversification) with an improvement on IR of 25$\%$ on average annually.

\item We have developed a systematic framework for asset allocation using  an embedded feature selection algorithm to identify features of relevance to the model. This improves the model's accuracy and allows for a more objective approach to portfolio construction in the sense that it should help to prevent biases in the feature selection process which is normally done by a financial expert. 

We used a FSHMM algorithm to select relevant features from a pool of well known factor indices and compared it with a HMM trained with the whole set of assets. Both models showed agreement on regime identification, with the model trained using only relevant features being more sensitive to periods of economic distress. 
\item We have tested both models using real, investable assets through MSCI USA enhanced factor indices. Portfolios constructed using information from the FSHMM trained with relevant features show a higher performance than the same portfolios constructed using a HMM trained with full set of features. 
\end{enumerate}

Possible extensions of the model for future work could be to include macroeconomic series in the HMM, where the embedded feature selection could potentially solve the problem of selecting relevant economic series, allowing for a more precise identification of economic cycles. This would be particularly interesting for other asset classes such as fixed income, but this is outside of the scope of this paper.

A drawback of using HMMs is that the number of latent states has to be known in advance, or selected through BIC, which is not always effective, or with a greedy approach choosing the model with higher performance. This could be addressed using an infinite HMM \citep{iHMM}.

\section*{Acknowledgement}
The authors are thankful to Sahil Kahn, David Hutchins and Andrew Chin for their valuable feedback on early results of this work. This work was supported by the European Union's Horizon 2020 research and innovation programme under the Marie Sklodowska-Curie Grant Agreement no. 675044 (\url{http://bigdatafinance.eu/}), Training for Big Data in Financial Research and Risk Management.

\appendix

\section{Feature saliency HMM}
\label{appendix:FS}
The FSHMM algorithm as developed by Adams, Beiling and Cogill has the following EM update steps (for simplicity we follow their notation):
\subsubsection*{E-Step}
\begin{eqnarray}
 \gamma_t(i) & = & P(x_t=i|y, \Lambda')  \label{eq:Estep01}\\
 \xi(i,j) & = & P(x_{t-1} = i, x_t = j| y, \Lambda')  \label{eq:Estep02}
\end{eqnarray}
With $\gamma_t(i)$ and $\xi(i,j)$ calculated with the forward-backward algorithm. The additional updates are:
\begin{eqnarray}
 e_{ilt} & = & \rho_l r(y_{lt}|\mu_{il},\sigma^2_{il})  \label{eq:Estep1}\\
 h_{ilt} & = & (1-\rho_l)q(y_{lt}|\epsilon_l, \tau^2_l)  \label{eq:Estep2}\\
 g_{ilt} & = & e_{ilt} + h_{ilt}  \label{eq:Estep3}\\
 u_{ilt} & = & \frac{\gamma_{it} e_{ilt}}{g_{ilt}}  \label{eq:Estep4}\\
 v_{ilt} & = & \gamma_{it} - u_{ilt} \label{eq:Estep5}
\end{eqnarray}

\subsubsection*{MAP M-step:}

\begin{eqnarray}
\pi_i & = & \frac{\gamma_0 (i) +\beta_i - 1}{\sum_{i=1}^I (\gamma_0(i) + \beta_i - 1)} \label{eq:Mstepi}\\
a_{ij} & = & \frac{\sum_{t=1}^T \xi_t(i,j) +\alpha_{ij}-1}{\sum_{j=1}^I(\sum_{t=1}^T \xi_t(i,j) + \alpha_{i,j -1})} \\
\mu_{il} & = & \frac{s_{il}^2 \sum_{t=0}^T u_{ilt} y_{lt} + \sigma^2_{il} m_{il}}{s_{il}^2 \sum_{t=0}^T u_{ilt} + \sigma^2_{il}}\\
\sigma^2_{il} & = & \frac{\sum_{t=0}^T u_{ilt} (y_{lt}-\mu_{il})^2 + 2\eta_{il}}{\sum_{t=0}^T u_{ilt} + 2 (\zeta_{il} + 1)}\\
\epsilon_l & = & \frac{c^2_l \sum_{t=0}^T(\sum_{i=1}^I v_{ilt}) y_{ilt} + \tau^2_l b_l}{c_l^2 \sum_{t=0}^T (\sum_{i=1}^I v_{ilt}) + \tau^2_l} \\
\tau^2_l & = & \frac{\sigma_{t=0}^T (\sum_{i=1}^I v_{ilt}) (y_{lt}-\epsilon_l)^2 + s\psi_l}{\sigma_{t=0}^T (\sum_{i=1}^I v_{ilt}) + 2(v_l + 1)} \\
\rho_l & = & \frac{\hat{T} - \sqrt[2]{\hat{T}^2 -4k_l (\sum_{t=0}^T \sum_{i=1}^I u_{ilt})}}{2k_l}
\label{eq:Mstepf}
\end{eqnarray}
where $\hat{T} = T + 1+k_l$.

Table \ref{table:rhos1} shows feature saliency of 5 relevant features and three irrelevant features generated with $\mathcal{N}(0,1)$ with different number of observations and number of hidden states. Table \ref{table:rhos2} shows the same but with 10 relevant features and 5 added series of noise, for different states and values of $k$ parameter.

\begin{table*}
    \caption{Feature saliency of five factor returns time series ($\rho_1$ to $\rho_5$) and three irrelevant series of random noise ($\rho_6$ to $\rho8$), all calculated with $k=50$. All irrelevant features have saliency below 0.25, and most of the financial series have saliency close to one, except $\rho_3$ that has a small saliency in most of the cases.}
    \ra{1.2}
    \centering
    \small
    \begin{tabular}{@{}llccccccccc@{}}\toprule
        \multicolumn{2}{l}{Case} & \phantom{i} &$\rho_1$ & $\rho_2$ & $\rho_3$ & $\rho_4$ & $\rho_5$ & $\bm{\rho_6}$ & $\bm{\rho_7}$ & $\bm{\rho_8}$\\
        \cmidrule{1-2} \cmidrule{4-11} 
    500 points & 2 states & & 0.990 & 0.971 & 0.307 & 0.987 & 0.966 & 0.141 & 0.042 & 0.047\\
    500 points & 3 states & & 0.991 & 0.990 & 0.264 & 0.987 & 0.988 & 0.171 & 0.035 & 0.071\\ 
    2000 points & 2 states & & 0.986 & 0.986 & 0.190 & 0.994 & 0.995 & 0.017 & 0.007 & 0.018\\
    2000 points & 3 states & & 0.996 & 0.997 & 0.123 & 0.996 & 0.996 & 0.066 & 0.202 & 0.033\\
        \bottomrule
    \end{tabular}
    \label{table:rhos1}
\end{table*}

\begin{table*}
    \caption{Feature saliency of ten factor returns time series ($\rho_1$ to $\rho_{10}$) and five irrelevant series of random noise ($\rho_{11}$ to $\rho_{15}$). With a small value of $k$ all irrelevant features are discarded and all relevant features have high saliency. With a larger k, noise features are discarded, but also financial features start being selected. All series have 2000 observations.}
    \ra{1.2}
    \centering
    \small
  \begin{tabular}{@{}llcccccccccccccccc@{}}\toprule
         \multicolumn{2}{l}{Case} & \phantom{i} &$\rho_1$ & $\rho_2$ & $\rho_3$ & $\rho_4$ & $\rho_5$ & $\rho_6$ & $\rho_7$ & $\rho_8$ & $\rho_9$ & $\rho_{10}$ & $\bm{\rho_{11}}$ & $\bm{\rho_{12}}$ & $\bm{\rho_{13}}$ & $\bm{\rho_{14}}$ & $\bm{\rho_{15}}$\\
        \cmidrule{1-2} \cmidrule{4-18} 
    $k = 50$ & 2 states & & 0.99 & 0.99 & 0.56 & 0.99 & 0.91 & 1.00 & 0.99 & 0.95 & 0.99 & 0.97 & 0.11 & 0.11 & 0.04 & 0.26 & 0.07\\
    $k = 50$ & 3 states & & 1.00 & 0.99 & 1.00 & 1.00 & 1.00 & 1.00 & 1.00 & 1.00 & 1.00 & 0.99 & 0.24 & 0.09 & 0.40 & 0.10 & 0.11\\
    $k = 200$ & 2 states & & 0.75 & 0.03 & 0.13 & 0.98 & 0.44 & 0.99 & 0.99 & 0.17 & 0.98 &  0.14 & 0.05 & 0.02 & 0.02 & 0.01 & 0.04 \\
    $k = 200$ & 3 states & & 1.00 & 0.37 & 0.08 & 0.99 & 0.55 & 1.00 & 1.00 & 0.13 & 0.99 & 0.22 & 0.04 & 0.17 & 0.04 & 0.05 & 0.03\\

        \bottomrule
    \end{tabular}
    \label{table:rhos2}
\end{table*}

\section{Portfolio description}
\label{appendix:port}
All portfolios constructed are long only, i.e. $w \geq 0$.
\begin{itemize}
\item {\bf Max return:} Given an estimated vector of means, it maximizes the return given a constrain that no asset can have a weight greater than $80\%$. 
\item {\bf Dyn:} If all estimated mean asset returns are positive, it weights the assets proportional to their mean, else, it equally weights them. 
\item {\bf Sharpe:} is a classic mean-variance portfolio that maximizes return given a set level of risk.
\item {\bf Risk parity:} focuses on the allocation of risk, each asset on the portfolio contributes the same risk as defined by 
\begin{equation*}
\frac{w_j(Vw)_j}{\sqrt{wVw'}}
\end{equation*}
where $V$ is the covariance matrix.
\item {\bf Max diversification} Maximizes the diversification ratio defined as:
\begin{equation*}
\frac{w'\Sigma}{\sqrt[2]{w'Vw}}
\end{equation*}
where $\Sigma$ is a vector of all asset volatility and $V$ is the covariance matrix.
\item {\bf Min Var:} finds the portfolio with minimum variance, defined by:
\begin{equation*}
w'Vw
\end{equation*}
where $V$ is the covariance matrix.
\end{itemize}


\pagebreak

\bibliography{sample}

\end{document}